\documentclass[journal]{IEEEtran}
\usepackage{graphicx}
\usepackage{caption}
\usepackage{amsmath}
\usepackage{bbm}
\usepackage{multirow}
\usepackage[numbers,sort&compress]{natbib}
\usepackage{subfigure}
\usepackage[hidelinks]{hyperref}
\usepackage[linesnumbered,ruled]{algorithm2e}
\usepackage{harpoon}
\usepackage{color}
\usepackage{amssymb}
\newtheorem{remark}{Remark}

\ifCLASSINFOpdf
\else
\fi
\usepackage{geometry}
\begin{document}
%\title{Task-oriented Communications via the VoI and AoI based Queue Ordering
%\thanks{This work was supported by Engineering and Physical Sciences Research Council (EPSRC), U.K., under Grant EP/W004348/1.}
%}

\title{Guaranteed Image Classification via Goal-oriented Joint Semantic Source and Channel Coding\\
}
%Guaranteed Classification via Goal-oriented Joint Semantic and Channel Coding
 
%想一下performance metric的具体term (classification probability or desired class probability)

\author{Wenchao Wu,
        Min Qiu,
        Yansha Deng,
        and Jinhong Yuan\\
        % <-this % stops a space
        \thanks{W. Wu and Y. Deng are with the Department of Engineering, King’s College London, Strand, London WC2R 2LS, U.K. M. Qiu is with Global College, Shanghai Jiao Tong University, Shanghai, China. J. Yuan is with School of Electrical Engineering and Telecommunications, University of New South Wales, Sydney, Australia (e-mail: wenchao.wu@kcl.ac.uk; min\_qiu@sjtu.edu.cn; yansha.deng@kcl.ac.uk; j.yuan@unsw.edu.au). (Corresponding author: Yansha Deng). The conference version has been submitted to IEEE ICC 2026.}
}

\maketitle
\pagestyle{empty}
\thispagestyle{empty}
\begin{abstract}
To enable critical applications such as remote diagnostics, image classification must be guaranteed under bandwidth constraints and unreliable wireless channels through joint source and channel coding (JSCC) design. However, most existing JSCC methods focus on minimizing image distortion, implicitly assuming that all image regions contribute equally to classification performance, thereby overlooking their varying importance for the task. In this paper, we propose a goal-oriented joint semantic source and channel coding (G-JSSCC) framework that applies \emph{various} levels of source coding compression and channel coding protection across image regions based on their semantic importance. Specifically, we design a semantic information extraction method that identifies and ranks various image regions based on their contributions to classification, where the contribution is measured by the shapely value from explainable artificial intelligence (AI). Based on that, we design a 
semantic source coding and a semantic channel coding method, which allocates higher-quality compression and stronger error protection to image regions of great semantic importance. In addition, we define a new metric, termed coding efficiency, to evaluate the effectiveness of the source and channel coding in the classification task. Simulations show that our proposed G-JSSCC framework improves classification probability by 2.70 times, reduces transmission cost by 38$\%$, and enhances coding efficiency by 5.91 times, compared to the benchmark scheme using uniform compression and an idealized channel code to uniformly protect the whole image.

\end{abstract}


% \fi
\IEEEpeerreviewmaketitle

\section{Introduction}
%介绍传统无线通信并引申到goal-oriented semantic communication
%为什么goal 是 image classification
%现在的 goal-oriented image classification都是显性的，比如deepJSCC, 把整张image看成整体，因此对一张图片用同样的channel coding
%然而，不同 part的image有不同的重要性，需要不容的channel coding保护
% 设计不同length的编码
%贡献
% 文章结构

%介绍传统无线通信并引申到goal-oriented semantic communication
\IEEEPARstart{T}{he} rapid development of machine learning \cite{ML_BG} and wireless communication technologies \cite{6G_BG} has facilitated the emergence of intelligent wireless systems capable of performing high-level cognitive tasks \cite{Background}, including image classification \cite{Img_Cla}, image recognition \cite{Img_Rec}, and image retrieval \cite{Img_Ret}. Among these tasks, image classification has received considerable attention due to its important role in mission-critical applications such as remote diagnostics \cite{Remo_diag}, autonomous driving \cite{Auto_Driving}, and industrial automation \cite{Ind_Aut}. However, in such safety-critical applications, erroneous decisions arising from misclassified images can be particularly severe, potentially endangering human lives and causing damage to the surrounding environment.

\par To address this challenge, accurate image classification is essential, where the classification probability must be guaranteed to exceed a predefined threshold under varying channel conditions. Since the communication resources is limited in practical, achieving such robustness necessitates the reliable transmission of images with minimal transmission cost. However, raw image data typically contains significant statistical redundancy \cite{Motivate_source_coding}, and practical wireless communication channels are inherently affected by impairments such as noise, interference, and fading \cite{Motivate_channel_coding}. Directly transmitting raw image data over noisy wireless channels not only incurs excessive transmission cost but also compromises reliability. To solve this problem, it is essential to employ source and channel coding \cite{Coding}. Source coding is introduced to reduce the amount of data required to represent an image, thus minimizing the number of bits that need to be transmitted over the wireless channel. This is commonly achieved following existing image compression standards, such as the joint photographic experts group (JPEG) \cite{JPEGOri} and JPEG 2000 \cite{JPEG2k}. Specifically, JPEG compresses images by applying a block-based discrete cosine transform (DCT) followed by quantization and encoding, but introduces some information loss. In contrast, JPEG 2000 applies a discrete wavelet transform (DWT) over the entire image with improved compression efficiency. Channel coding, on the other hand, 
aims to detect and correct errors by introducing structured redundancy to image bit stream. In the fifth generation (5G) new radio (NR) specification \cite{3GPP}, standard channel coding schemes include low-density parity-check (LDPC) codes \cite{LDPC} and polar codes \cite{Polar}, which protect the image transmission by adding redundant bits, using sparse linear block codes and polar transform. Existing communication standards treat source coding and channel coding as separate and independent modules, which poses risks of failing to meet the desired classification performance requirements under dynamic channel conditions \cite{SSCC}.

\par Many existing studies focused on jointly optimizing the source and channel coding, referred to as joint source and channel coding (JSCC) \cite{JSCC_BG}. Traditional JSCC designs rely on mathematical models of source and channel characteristics, such as Markov and Gaussian distributions, to perform the joint design of source and channel codewords \cite{JSCC_1,JSCC_2,JSCC_3}. More recently, deep learning-based JSCC schemes (DeepJSCC) \cite{DeepJSCC} have been proposed to solve joint source and channel coding via end-to-end deep learning, by replacing conventional encoders and decoders with convolutional neural networks (CNN). This compression and protection of the image is achieved by jointly optimizing the CNNs' parameters to minimize the average distortion between the input image and its reconstruction or a rate distortion loss function that balances coding rate reduction with image distortion minimization \cite{DeepJSCC_1,DeepJSCC_2,DeepJSCC_3}. Until now, existing JSCC designs compress and protect the image by treating it as a whole entity, implicitly assuming that all image regions contribute equally to the classification performance. However, in practice, different regions of an image contribute unequally to classification, and some regions may negatively impact the classification performance. Allocating higher compression quality factor or stronger protection to less relevant or detrimental image regions can reduce the overall image classification probability. Consequently, applying uniform compression and protection across the entire image may not only affect optimal classification probability but also introduce unnecessary transmission cost that do not improve, and may even degrade the classification probability.

\par To overcome these limitations, higher compression quality factor and stronger protection should be allocated to more important image regions. Prior work, such as \cite{UEP,UEP_1,UEP_2}, has explored unequal error protection (UEP) in data compression and transmission by allocating greater redundancy to bits deemed more important, assuming the importance is either predefined or related to the image reconstruction. However, in the context of image classification, importance is inherently tied to the classification task itself, and the question of how to identify, quantify, and rank the significance of each image region optimally has not been fully addressed. The semantic information extraction, a method introduced in goal-oriented semantic communication, can quantify the importance of transmitted information with respect to the effectiveness metric of diverse tasks and applications \cite{Goal_semantic_communication}. For example, to achieve accurate avatar moving of the AR application, the importance of each skeleton in an avatar-centric displaying task can be evaluated by considering its number of connections and distances from other connected skeletons in \cite{Goal_seman_exmple1}. To minimize the tracking error in the UAV waypoint transmission task, the importance of each control and command data can be measured in \cite{Goal_seman_exmple2} as a function of its age of information (AoI) and value of information (VoI). To minimize the communication load under the digital twin reconstruction error constraints in the robot arm reconstruction task, the importance of each reconstruction message has been quantified in \cite{Goal_seman_exmple3} based on the current movement of the robot. By leveraging the semantic information extraction method, it becomes feasible to obtain the importance of different image regions with respect to the classification task. This enables to apply appropriate levels of source coding compression and channel coding protection to different image regions with varying sizes.

\par Although existing source coding standards are capable of accommodating variability in information length across image regions, standard channel coding schemes impose strict constraints on codeword lengths and allowable code rate ranges \cite{3GPP}. These limitations pose significant challenges of applying heterogeneous levels of protection to image regions of different sizes. To simultaneously support a broader range of codeword lengths and rates with near-optimal performance, a universal coding scheme is required. Generalized spatially coupled parallel concatenated codes (GSC-PCCs) \cite{9851473} are promising candidates for this purpose due to two key advantages: 1) it has been proved that a single GSC-PCC ensemble suffices to achieve capacity for any rate in $\left(0,1\right)$ under random puncturing, eliminating the need to explicitly optimize the code structure for a particular code rate as in the conventional block coding; 2) GSC-PCCs have flexible structures that do not impose constraints on information and codeword lengths, and can be decoded using sliding window decoding with manageable latency.

\par Motivated by this, we propose a novel goal-oriented joint semantic source and channel coding (G-JSSCC) framework for transmitting images over wireless channels to achieve classification task, which integrates importance in advanced source and channel coding to guarantee the minimum image classification probability. Unlike existing source and channel coding schemes that uniformly compress and encode the entire image, our proposed framework designs a semantic information extraction method to identify and rank the importance of various image regions based on their contributions to the image classification task. As the task is typically executed by an artificial intelligence (AI)-based classification model, the importance of different image regions can be quantified by integrating explainable AI techniques (e.g., the shapley value) into the semantic information extraction method, which provides interpretability to explain how different image regions influence the classification probability. We also design a G-JSSCC approach to allow image regions of greater semantic importance to receiver higher quality compression and stronger error protection. Our main contributions are summarized as follows:
\begin{figure*}[htbp]
    \centering
    \subfigure[Traditional communication framework.]{
    \includegraphics[width=0.9\linewidth]{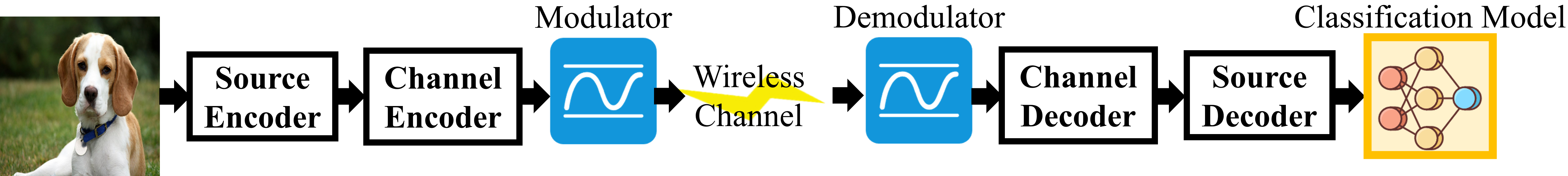}
    }
    \subfigure[Proposed G-JSSCC framework.]{
    \includegraphics[width=0.9\linewidth]{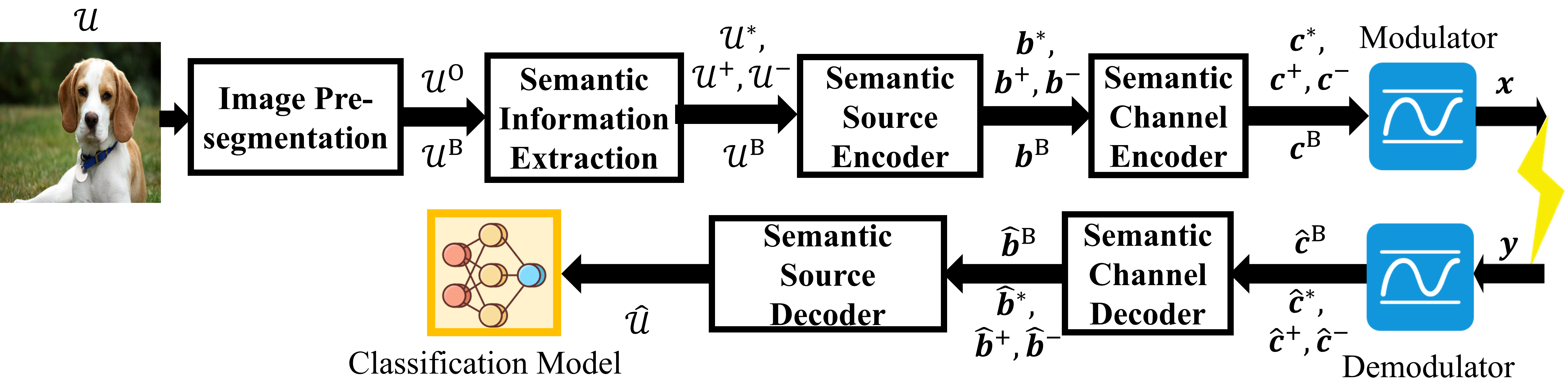}
    }
    \caption{Frameworks.}
    \label{Fig:Frameworks}
\end{figure*}
\begin{itemize}
    \item We propose a G-JSSCC framework that satisfies the minimum image classification probability threshold under various channel conditions. This framework includes image pre-segmentation, semantic information extraction, semantic source coding, and semantic channel coding. The image pre-segmentation method first separates the input image into the background and object, with the object further divided into multiple regions. Then, for a target classification probability threshold, the semantic information extraction method resegments the object. After that, the semantic source coding and semantic channel coding methods apply unequal level of compression and protection to various image regions, respectively.
    \item We design a semantic information extraction method to resegment the object based on each object region's importance to classification. Unlike existing works that focus on end-to-end image distortion, we explicitly consider the joint effects of source and channel coding in classification. This method includes shapley value-based segmentation and semantic information-based importance ranking. First, the proposed segmentation method quantifies each object region's importance in an \emph{interpretable} manner using shapley value calculation from explainable AI, which reflects the improvement in classification probability when this region experiences higher level of source coding compression and channel coding protection. Then, the semantic information of the image is extracted by aggregating object regions with higher shapley values into the most important superpixel. After that, by integrating semantic information in the calculation of shapley value, the semantic information-based importance ranking method categorizes the remaining object regions as positive and negative superpixels.
    \item We design a semantic source coding and a semantic channel coding method. To guarantee the classification probability with minimum transmission cost, the proposed G-JSSCC operates by optimizing the coding parameters rather than relying on any specific codec. Specifically, the proposed scheme is flexible with respect to off-the-shelf source coding (e.g., JPEG or JPEG 2000) and channel coding (e.g., GSC-PCCs), and can be readily configured based on the optimized coding parameters. Essentially, the semantic source coding module assigns higher quality-factor settings to image regions of greater semantic importance. Subsequently, the semantic channel coding module allocates lower channel code rate for these compressed image regions to achieve a lower channel bit error rate (BER).
    \item We define a new metric to evaluate the effectiveness of source and channel coding in the classification task, namely, coding efficiency. It is calculated as the product of the classification probability and the code rate, quantifying how effectively the source and channel coding enhances the classification probability.
    \item We conduct experiments under both good and poor channel quality conditions. With poor channel quality, the entire image is transmitted, whereas with good channel quality, we consider transmission \emph{without the background} to further reduce the transmission cost. Compared to the traditional communication framework that uniformly compresses and protects the whole image, our proposed G-JSSCC framework could improve classification probability by 2.70 times, reduce the transmission cost by 38$\%$, and enhance coding efficiency by 5.91 times.
\end{itemize}
%\begin{figure*}[h]
%    \centering
%    \includegraphics[width=\textwidth]{Figure/Proposed Framework.png}
%    \caption{Overview of the proposed G-JSemSCC framework.}
 %   \label{Fig:Framework}
%\end{figure*}

\par The rest of the paper is organized as follows: In Section $\mathrm{\uppercase\expandafter{\romannumeral2}}$, we present the system model and problem formulation. Section $\mathrm{\uppercase\expandafter{\romannumeral3}}$ introduces the design of our proposed G-JSSCC framework. In Section $\mathrm{\uppercase\expandafter{\romannumeral4}}$, the simulation results is presented. Finally, Section $\mathrm{\uppercase\expandafter{\romannumeral5}}$ concludes this paper.
\section{System Model \& Problem Formation}
In this section, we first present the traditional communication framework for image classification. Then, we introduce our novel G-JSSCC framework. Finally, we present our performance metrics and the problem formulation associated with our proposed framework.
\subsection{System Model}
\subsubsection{Traditional communication framework} 
%不带notation
The traditional communication framework is shown in Fig. \ref{Fig:Frameworks} (a), where the transmitter consists of a source encoder, a channel encoder, and a modulator, and the receiver includes a demodulator, a channel decoder, a source decoder, and a classification model. At the transmitter, the source encoder uniformly compresses the input image into a binary bit stream. The channel decoder then applies uniform protection to the bit stream and encodes it to a codeword. Finally, the modulator converts the codeword into a symbol sequence and transmits it across a wireless channel. At the receiver, the received symbol sequence is first demodulated by the demodulator, followed by the channel decoder to recover the transmitted bit stream. The recovered bit stream then undergoes the source decoder to reconstruct the input image, which is passed to the classification model to perform image classification.

\subsubsection{Overview of our proposed G-JSSCC framework}
%Semantic informtion extraction
%Semantic Coding
%Semantic Decoding
% Transmitter有什么， receiver有什么
% 然后按照过程介绍
%只讲输入输出
As shown in Fig. \ref{Fig:Frameworks} (b), to guarantee classification performance with transmission cost, our proposed G-JSSCC framework integrates joint semantic source and channel coding \footnote{In our work, the joint aspect of semantic, source coding, and channel coding comes from their mutual dependence. Specifically, source coding is based on semantic information and region importance, which relate to the different protection levels that channel coding can provide. Channel coding is also designed based on region importance. Thus, they are co-designed to meet the same goal-oriented performance requirement, rather than being optimized separately.}, which employs a higher compression quality factor and stronger protection to more important image regions. The transmitter includes image pre-segmentation, semantic information extraction, a semantic source encoder, a semantic channel encoder, and a modulator. The receiver comprises a demodulator, a semantic channel decoder, a semantic source decoder, and a classification model.

\par At the transmitter, the image pre-segmentation method separates input image $\mathcal{U}=\{\mathcal{U}^{\mathrm{B}},\mathcal{U}^{\mathrm{O}}\}$ into background $\mathcal{U}^{\mathrm{B}}$ and object $\mathcal{U}^{\mathrm{O}}$, with $\mathcal{U}^{\mathrm{O}}$ further divided into multiple regions. Then, the semantic information extraction approach resegments $\mathcal{U}^{\mathrm{O}}=\{\mathcal{U}^{*},\mathcal{U}^{+},\mathcal{U}^{-}\}$ into the most important superpixel $\mathcal{U}^{*}$ that solely guarantees classification after source and channel coding, positive superpixels $\mathcal{U}^{+}$ that enhance classification after source and channel coding, and negative superpixels $\mathcal{U}^{-}$ that degrade classification after source and channel coding. After that, the semantic source encoder converts these image regions $\mathcal{U}^{\mathrm{B}}$, $\mathcal{U}^{\mathrm{*}}$, $\mathcal{U}^{\mathrm{+}}$, and $\mathcal{U}^{\mathrm{-}}$ to bit streams $\boldsymbol{b}^{\mathrm{B}}$, $\boldsymbol{b}^{*}$, $\boldsymbol{b}^{+}$, and $\boldsymbol{b}^{-}$, with different levels of compression quality factor. Furthermore, the semantic channel encoder applies various levels of protection to various bit streams and generates codewords $\boldsymbol{c}^{\mathrm{B}}$, $\boldsymbol{c}^{*}$, $\boldsymbol{c}^{+}$, and $\boldsymbol{c}^{-}$. Finally, the modulator modulates these codewords into symbol sequences $\boldsymbol{x}^{\mathrm{B}}$, $\boldsymbol{x}^{*}$, $\boldsymbol{x}^{+}$, and $\boldsymbol{x}^{-}$, which are concatenated to $\boldsymbol{x}$ and transmitted over the wireless channel.

\par At the receiver, the demodulator demodulates the received symbol sequence $\boldsymbol{y}$ and obtain the soft estimation of each coded bit, i.e., the log-likelihood ratio (LLR). Then, the semantic channel decoder decodes the LLR and output the estimated bit streams $\hat{\boldsymbol{b}}^{\mathrm{B}}$, $\hat{\boldsymbol{b}}^{*}$, $\hat{\boldsymbol{b}}^{+}$,and $\hat{\boldsymbol{b}}^{-}$. From these estimated bit streams, the semantic source decoder obtains the corrupted image $\hat{\mathcal{U}}$, which is finally fed into the classification model to perform the classification task.

\subsection{Wireless Channel Model}
The wireless channel model is characterized by the widely adopted quasi-static Rayleigh fading channel. Under this channel model, the received signal is given by
\begin{equation}
    \boldsymbol{y}=h\boldsymbol{x}+\boldsymbol{z},
    \label{eq:Channel}
\end{equation}
where the transmitted signals $\boldsymbol{x}$ of blocklength $N$ are subject to a power constraint $\frac{1}{N}\mathbb{E}[\|\boldsymbol{x}\|^2] \leq P$, $h\sim \mathcal{CN}\left(0, \sigma_{h}^{2}\right)$ is a complex normal random variable representing the channel coefficient, and each element of the noise vector $\boldsymbol{z}$ follows an independent identically distributed (i.i.d.) complex Gaussian distribution, i.e., $z\sim \mathcal{CN}(0, \sigma^2_z)$.

\par In this paper, the binary phase-shift keying (BPSK) modulation is considered for simplicity, whereas other high-order modulations can be easily adapted. Based on the channel model in Eq. \eqref{eq:Channel}, the uncoded BPSK signal has an average error probability of
\begin{equation}\label{eq:BPSK_BER}
\epsilon_c = Q\left(\sqrt{\frac{2P|h|^2}{\sigma^2_z}}\right),
\end{equation}
where $Q\left(\cdot\right)$ is the Q-function, defined as
\begin{equation}
Q(x) = \frac{1}{\sqrt{2\pi}}\int_{x}^{\infty}e^{-\frac{t^{2}}{2}}\mathrm{d}t.
\end{equation}

\subsection{Performance Metrics}
\subsubsection{Classification Probability of Specific Class}
An image classification model is always trained to identify $C$ classes, indexed by $\mathcal{C}=\{1,2,..,C\}$. Given the recovered image $\hat{\mathcal{U}}$, the model outputs probability distribution $\boldsymbol{p}_{\hat{\mathcal{U}}}=\{p^{1}_{\hat{\mathcal{U}}},...p_{\hat{\mathcal{U}}}^{c},...,p_{\hat{\mathcal{U}}}^{C}\}$, where $p_{\hat{\mathcal{U}}}^{c}$ represents the probability of recognizing $\hat{\mathcal{U}}$ as class $c$, satisfying $\sum_{c\in\mathcal{C}}p_{\hat{\mathcal{U}}}^{c}=1$. As the objective is to classify $\hat{\mathcal{U}}$ as a specific class $D\in\mathcal{C}$, rather than considering the entire probability distribution $\boldsymbol{p}_{\hat{\mathcal{U}}}$, the focus is on the classification probability of the specific class $p_{\hat{\mathcal{U}}}^{D}$. To ensure reliable classification, $p_{\hat{\mathcal{U}}}^{D}$ must exceed a certain threshold, represented as $p_{\mathrm{th}}$.
\subsubsection{Coding Efficiency}  We denote the length of the raw image bit sequence as $K$ and the lengths of the final transmitted sequence as $N$. Accordingly, we can calculate \emph{code rate} as
\begin{equation}
    R=  \frac{K}{N}.
    \label{eq:JSemCC Rate}
\end{equation}
A lower $R$ implies more transmission cost. To measure how effectively the source and channel coding scheme influences the classification probability of a specific class $p^{D}_{\hat{\mathcal{U}}}$, a new performance metric, named \emph{coding efficiency}, is defined as
\begin{equation}
e^{D}_{\hat{\mathcal{U}}} \triangleq p^{D}_{\hat{\mathcal{U}}}\times R.
\end{equation}
A higher $e^{D}_{\hat{\mathcal{U}}}$ indicates a more efficient source and channel coding scheme that achieves higher classification performance with fewer transmitted costs.

\subsection{Problem Formulation}
The proposed G-JSSCC framework aims to guarantee $p^{D}_{\hat{\mathcal{U}}}$ with minimum transmission cost. Since transmission cost is directly associated with the code rate $R$, the optimization problem is formulated as minimizing $R$ under the constraint $p^{D}_{\hat{\mathcal{U}}}>p_{\mathrm{th}}$. The corresponding objective function is expressed as
\begin{equation}
\begin{aligned}
    &\mathcal{P}1: \max R
     \\
     &s.t.\ \ p^{D}_{\hat{\mathcal{U}}}>p_{\mathrm{th}}.\\
\end{aligned}
\end{equation}

\section{The Proposed G-JSSCC framework} 
In this section, we present the design principles for our proposed G-JSSCC framework in detail, including image pre-segmentation, semantic information extraction, and semantic coding design. 
\subsection{Image Pre-segmentation}
\begin{figure}[htbp]
    \centering
    \includegraphics[width=\linewidth]{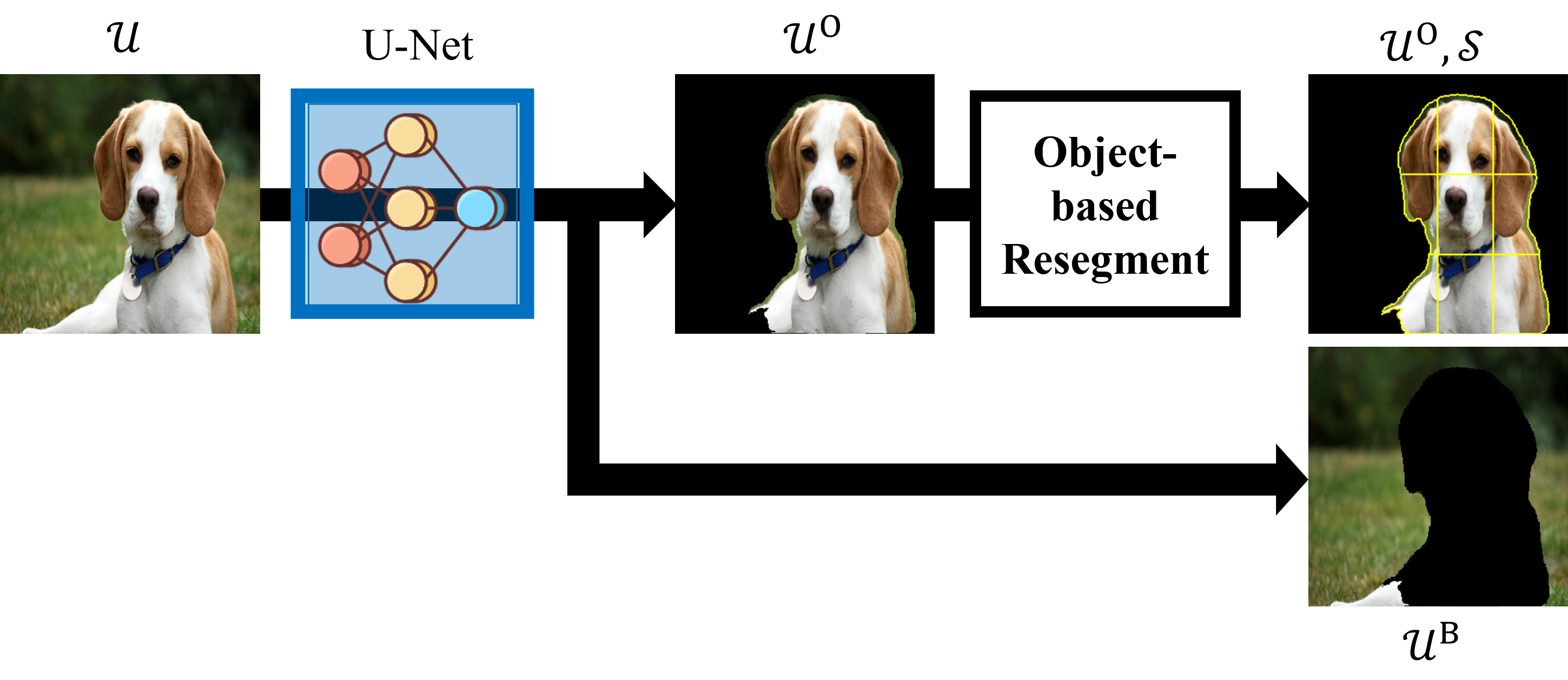}
    \caption{Image pre-segmentation.}
    \label{Fig:Image Pre-seg}
\end{figure}
An image $\mathcal{U}$ typically consists of the background and the object. Since the background generally provides no or negative contribution to classify the specific class, the focus is on segmenting only the object. This is accomplished through the designed image pre-segmentation process, as shown in Fig. \ref{Fig:Image Pre-seg}. First, the object $\mathcal{U}^{\mathrm{O}}$ and the background $\mathcal{U}^{\mathrm{B}}$ are separated using U-Net, a widely used CNN architecture designed for image semantic segmentation, outputting the object's boundary within the image. 
\begin{figure}[htbp]
    \centering
    \includegraphics[width=\linewidth]{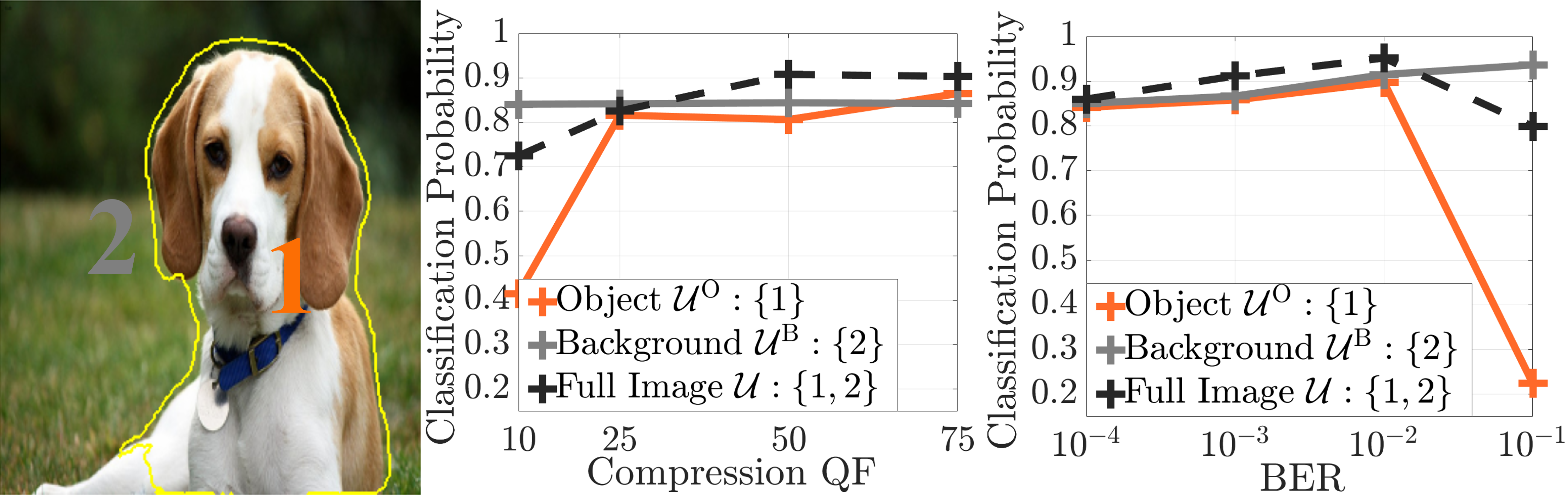}
    \caption{An example of preliminary simulations.}
    \label{Fig:Pre_Sim}
\end{figure}
\par Based on this, we have conducted preliminary simulations \footnote{The preliminary simulation have been conducted in multiple images, here we present the result of one image to clearly present the results and conclusions.} to examine how the background $\mathcal{U}^{\mathrm{B}}$ and the object $\mathcal{U}^{\mathrm{O}}$ affect the classification probability under different compression quality factors and channel qualities, respectively. In both JPEG and JPEG 2000 standards, the compression quality factor can be explicitly controlled, where a higher compression quality factor corresponds to allocating more bits for compression, resulting in improved reconstruction quality. Channel quality is quantified by BER (e.g., $10^{-1}$ denotes 1 error per 10 bits), where lower BER indicates better channel quality. One example is shown in Fig. \ref{Fig:Pre_Sim}, with $\mathcal{U}^{\mathrm{O}}$ (orange 1) and $\mathcal{U}^{\mathrm{B}}$ (gray 2) on the left side, the middle side plots classification probability $p^{D}_{\hat{\mathcal{U}}}$ comparisons under three cases:
\begin{itemize}
    \item \textbf{Orange line}: $\mathcal{U}^{\mathrm{O}}$ is compressed over varying quality factor, while $\mathcal{U}^{\mathrm{B}}$ is compressed with the best quality factor (e.g., 100).
    \item \textbf{Gray line}: $\mathcal{U}^{\mathrm{B}}$ is compressed over with varying quality factor, while $\mathcal{U}^{\mathrm{O}}$ is compressed with the best quality factor (e.g., 100).
    \item \textbf{Orange line}: The full image $\mathcal{U}$ is compressed with varying quality factor.
\end{itemize}
Contrary to common sense that higher compression quality factor consistently improves classification, $p^{D}_{\hat{\mathcal{U}}}$ for the full image first increases and then remains stable with improved compression quality factor. When only $\mathcal{U}^{\mathrm{B}}$ is subjected to lower compression quality factor, $p^{D}_{\hat{\mathcal{U}}}$ remains stable, implying that compression of $\mathcal{U}^{\mathrm{B}}$ has almost no effect on classification. In contrast, when only $\mathcal{U}^{\mathrm{O}}$ experiences reduced compression quality factor, $p^{D}_{\hat{\mathcal{U}}}$ decreases, indicating that the compression quality factor of the object significantly influences classification. This counterintuitive observation leads to an important conclusion.
\begin{remark}
    \textit{In the image classification task, different image regions require different compression quality factor, with some image regions (e.g. background) able to experience low compression quality factor.}
    \label{Rem: 1}
\end{remark}
\par Similarly, the right side of Fig. \ref{Fig:Pre_Sim} plots classification probability $p^{D}_{\hat{\mathcal{U}}}$ comparisons in three cases:
\begin{itemize}
    \item \textbf{Orange line}: $\mathcal{U}^{\mathrm{O}}$ is transmitted over channels with varying quality to BER while $\mathcal{U}^{\mathrm{B}}$ is transmitted error-free.
    \item \textbf{Gray line}: $\mathcal{U}^{\mathrm{B}}$ is transmitted over channels with varying quality to BER while $\mathcal{U}^{\mathrm{O}}$ is ideal transmitted.
    \item \textbf{Orange line}:The full image is transmitted over channels with varying quality to BER.
\end{itemize}
Contrary to common sense that higher channel quality always improves classification, $p^{D}_{\hat{\mathcal{U}}}$ for the full image first increases and then decreases as channel quality improves. When only $\mathcal{U}^{\mathrm{B}}$ experiences a poorer quality channel, $p^{D}_{\hat{\mathcal{U}}}$ increases, implying that $\mathcal{U}^{\mathrm{B}}$ can hinder classification with stronger channel coding protection. Interestingly, when only $\mathcal{U}^{\mathrm{O}}$ experiences a poorer quality channel, $p^{D}_{\hat{\mathcal{U}}}$ initially improves and then declines, indicating that some parts of the object aid classification, while others harm classification with stronger channel coding protection. This counterintuitive observation leads to another important conclusion.
\begin{remark}
    \textit{In the image classification task, different image regions require different levels of protection, with some image regions (e.g., background) do not need any protection.}
    \label{Rem: 1}
\end{remark}

\par Motivated by this insight, our aim is to quantify the importance of each image region and adapt the coding strategies accordingly. Specifically, more important image regions are targeted for higher compression quality factor and stronger protection. Since certain parts of the object $\mathcal{U}^{\mathrm{O}}$ hinder the classification, we further segment $\mathcal{U}^{\mathrm{O}}$ to assess the importance of each part. Specifically, we identify the minimum bounding box enclosing the object and uniformly divide it into multiple rectangles. The object information located in different rectangles is defined as different object regions, indexed by $s\in\mathcal{S}=\{1,2,..,S\}$, forming object pre-segmentation $\mathcal{U}^{\mathrm{O}}=\left[\mathcal{U}_{1},\ldots,\mathcal{U}_{S}\right]$.

\subsection{Semantic Information Extraction}
\begin{figure}[htbp]
    \centering
    \includegraphics[width=\linewidth]{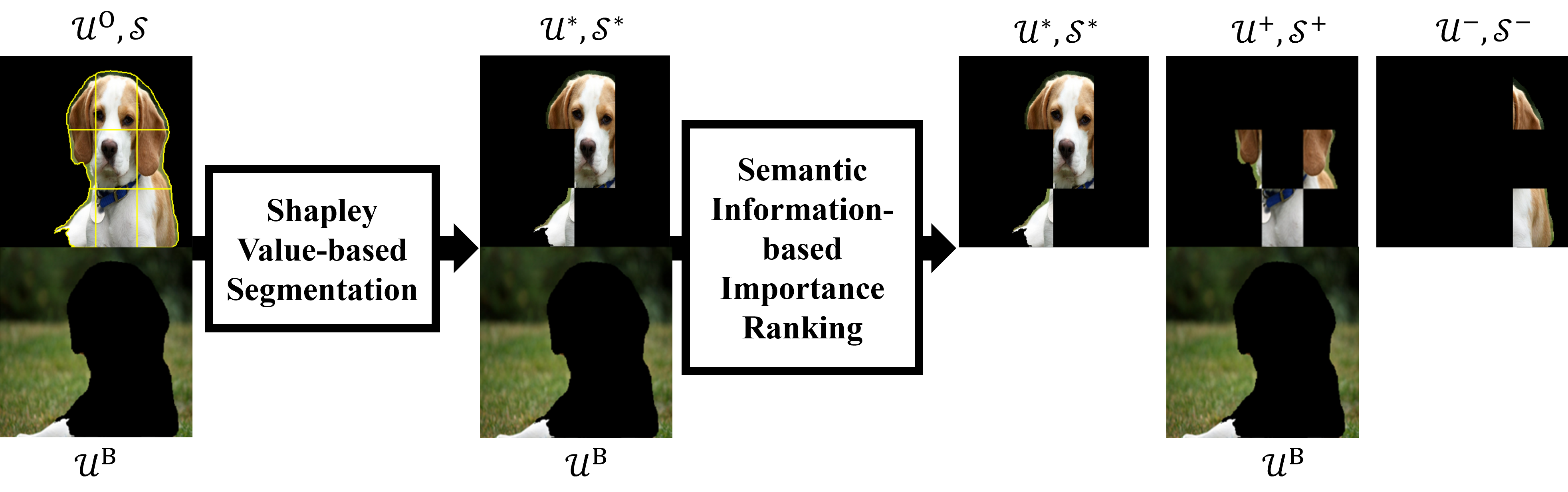}
    \caption{Semantic information extraction.}
    \label{Fig:Seman extract}
\end{figure}

Based on the observation in Fig. \ref{Fig:Pre_Sim}, the object $\mathcal{U}^{\mathrm{O}}$ could always be segmented into three parts:
\begin{itemize}
    \item \textbf{The most important superpixel} $\mathcal{U}^{*}$: It is obtained by aggregating several object regions with indexes in $\mathcal{S}^{*}$, which alone is sufficient to guarantee $p^{D}_{\hat{\mathcal{U}}}>p_{\mathrm{th}}$ when compressed in a higher quality factor and protected by stronger channel coding.
    \item \textbf{Positive superpixels} $\mathcal{U}^{+}$: The remaining object regions that enhance $p^{D}_{\hat{\mathcal{U}}}$ when compressed in a higher quality factor and protected by stronger channel coding, with indexes in $\mathcal{S}^{+}$.
    \item \textbf{Negative superpixels} $\mathcal{U}^{-}$: The remaining object regions that degrade $p^{D}_{\hat{\mathcal{U}}}$ when compressed in a higher quality factor and protected by stronger channel coding, with indexes in $\mathcal{S}^{-}$. 
\end{itemize}
As applying higher compression quality factor and stronger channel coding protection to more image regions increases the overall transmission cost, it is desirable to minimize the number of regions receiving such enhanced treatment. Since the most important superpixel $\mathcal{U}^{*}$ alone can guarantee $p^{D}_{\hat{\mathcal{U}}}>p_{\mathrm{th}}$ when compressed in a higher quality factor and protected by channel coding, it is sufficient to only apply such enhanced treatment to $\mathcal{U}^{*}$. Thus, the objective is to identify $\mathcal{U}^{*}$ by aggregating a minimum number of object regions. As a result, given a wireless channel BER $\epsilon_{c}$, a target BER after channel coding $\epsilon_{t}$, a basic compression quality factor $q_{b}$, and a target compression quality factor $q_{t}$, the problem $\mathcal{P}1$ could be convert to 
\begin{equation}
\begin{aligned}
    &\mathcal{P}2: \min |\mathcal{S}^{*}|
     \\
     &\mathrm{s.t.}\ \
    \Psi\left(\mathcal{U}^{*},\boldsymbol{b}^{*}\right)=q_{t},\\
     &\quad\ \ \  \Psi\left(\mathcal{U}^{\mathrm{B}},\boldsymbol{b}^{\mathrm{B}}\right)=\Psi\left(\mathcal{U}^{+},\boldsymbol{b}^{+}\right)=\Psi\left(\mathcal{U}^{-},\boldsymbol{b}^{-}\right)=q_{b},\\
     &\quad\ \ \ \mathrm{P_b}\left(\hat{\boldsymbol{b}}^{*},\boldsymbol{b}^{*}\right)=\epsilon_{t},\\
     &\quad\ \ \  \mathrm{P_b}\left(\hat{\boldsymbol{b}}^{\mathrm{B}},\boldsymbol{b}^{\mathrm{B}}\right)=\mathrm{P_b}\left(\hat{\boldsymbol{b}}^{+},\boldsymbol{b}^{+}\right)=\mathrm{P_b}\left(\hat{\boldsymbol{b}}^{-},\boldsymbol{b}^{-}\right)=\epsilon_{c},\\
     &\quad\ \ \ p^{D}_{\hat{\mathcal{U}}}>p_{\mathrm{th}},\\
\end{aligned}
\end{equation}
where $|\mathcal{S}^{*}|$ is the number of elements in set $\mathcal{S}^{*}$, indicating the number of object regions to form the most important superpixel $\mathcal{U}^{*}$. The function $\Psi\left(\mathcal{U},\boldsymbol{b}\right)=q$ is the source coding to compress the image $\mathcal{U}$ to the bit stream $\boldsymbol{b}$ with the quality factor $q$. The first constraint in $\mathcal{P}2$ indicates $\mathcal{U}^{*}$ is compressed by the target quality factor $q_{t}$, while the second constraint states that the remaining image regions are compressed by the basic quality factor $q_{b}$. The function $\mathrm{P_b}$ defines the BER associated with the bit stream of a particular image region. The third constraint in $\mathcal{P}2$ indicates the bit stream $\boldsymbol{b}^{*}$ of $\mathcal{U}^{*}$ is protected by channel coding, achieving target BER $\epsilon_{t}$, while the fourth constraint states that the bit streams of remaining image regions are not protected, experiencing wireless channel BER $\epsilon_{c}$. The parameters $\left(q_{b},q_{t},\epsilon_{c},\epsilon_{t}\right)$ determine the end-to-end image distortion, which in turn influences the classification probability $p^{D}_{\hat{\mathcal{U}}}$. Unlike existing works that focus on end-to-end image distortion in the problem formulation, we explicitly consider the joint impacts of target compression quality factor $q_{t}$ and target BER after channel coding $\epsilon_{t}$ on the classification probability $p^{D}_{\hat{\mathcal{U}}}$. To solve problem $\mathcal{P}2$, we design a semantic information extraction method, as shown in Fig. \ref{Fig:Seman extract}, including shapley value-based segmentation and semantic information-based importance ranking. 
\subsubsection{Shapley value-based segmentation}
To extract the most important superpixel $\mathcal{U}^{*}$ as the semantic information by aggregating several object regions, the contribution of each object region $s$ to the classification of the specific class $D$ must be quantified. This is achieved using the shapley value, a concept from cooperative game theory that has been adopted in explainable AI to assign fair attribution scores to individual contributors in a multi-agent system. For the index set $\mathcal{S}=\{1,\ldots,s,\ldots,S\}$, let $\mathcal{P}(\mathcal{S})$ denote the power set \cite{PowerSet} of $\mathcal{S}$ that has all subsets of $\mathcal{S}$, including the empty set $\emptyset$. We define the set $\mathcal{A}_{s}=\{ \mathcal{P} \in \mathcal{P}(\mathcal{S})\big| s \notin \mathcal{P}\}$ as all subsets of $\mathcal{S}$ that exclude the index $s$. To quantify the contribution of the object region $s$, we need to obtain how object region $s$ influences the classification probability $p^{D}_{\hat{\mathcal{U}}}$ after source and channel coding among all cases. Specifically, for the input image $\mathcal{U}$ and the subset $\mathcal{A}\in\mathcal{A}_{s}$, let $\mathrm{U}\left(\mathcal{A}\right)$ denote the reconstructed image, where the object regions with indexes in $\mathcal{A}$ experience target compression quality factor $q_{t}$ and are protected by channel coding to achieve a target BER $\epsilon_{t}$, while the remaining image regions experience basic compression quality factor $q_{b}$ and the channel BER $\epsilon_{c}$ without channel coding protection. Under this setup, the shapley value $v_{s}$ of the object region $s$ under $\left(q_{t},q_{b},\epsilon_{t}, \epsilon_{c}\right)$ is calculated as
\begin{equation}    v_{s}=\sum_{\mathcal{A}\in\mathcal{A}_{s}}\frac{|\mathcal{A}|!\left(|\mathcal{S}|-|\mathcal{A}|-1\right)!}{|\mathcal{S}|!}\left[p^{D}_{\mathrm{U}\left(\mathcal{A}\cup\{s\}\right)}-p^{D}_{\mathrm{U}\left(\mathcal{A}\right)}\right],
    \label{eq:SHAP}
\end{equation}
where $|\mathcal{A}|$ and $|\mathcal{S}|$ is the number of object region indexes in the subset $\mathcal{A}$ and set $\mathcal{S}$, $!$ is the factorial symbol, and $\mathcal{A}\cup\{s\}$ is the union of $\mathcal{A}$ and $\{s\}$. 
\begin{figure}[htbp]
    \centering
    \includegraphics[width=\linewidth]{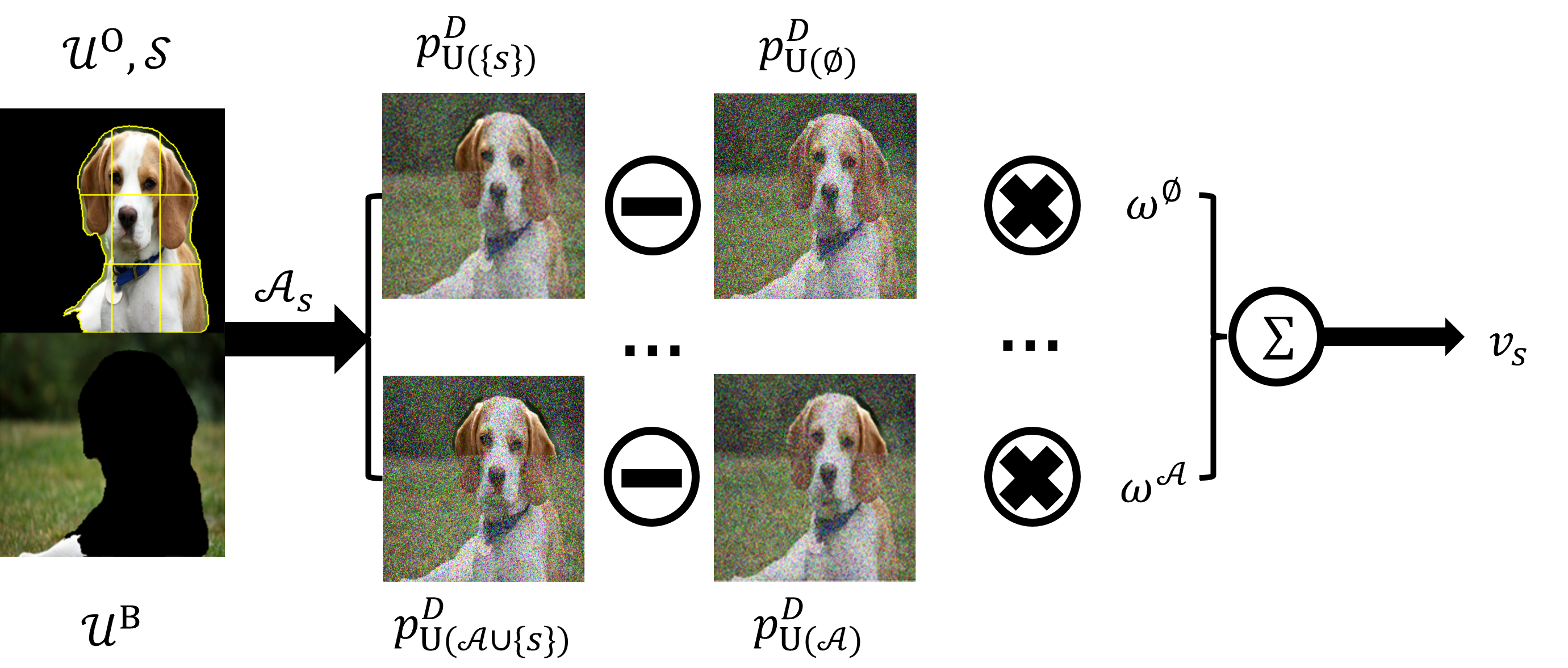}
    \caption{Shapley value calculation.}
    \label{Fig:Shapley calculation}
\end{figure}
\begin{algorithm}[h]
    \caption{Shapley value-based segmentation}
    \KwIn{$\mathcal{S}$, $\mathcal{U}=[\mathcal{U}^{\mathrm{B}},\mathcal{U}^{\mathrm{O}}]$, $p_{\mathrm{th}}$, $q_t$, $q_b$, $\epsilon_{c}, \epsilon_{t}$.} 
    Initialize $\mathcal{S}_{1}=\mathcal{S}$, $l=1$.\\
    Load classification model.\\
    \While{True}
    {
        \For{$s_{l}\in\mathcal{S}_l$}
        {
            $\mathcal{A}_{s_{l}}=\{\mathcal{A}\subseteq\mathcal{S}_{l}|s_{l}\notin\mathcal{A}\}$.\\
            $v_{s_{l}}^{l}=0$\\
            \For{$\mathcal{A}\in\mathcal{A}_{s_{l}}$}
            {
                Obtain $p^{D}_{\mathrm{U}\left(\mathcal{A}\cup\{s_{l}\}\right)}$ and $p^{D}_{\mathrm{U}\left(\mathcal{A}\right)}$ from classification model.\\
                Calculate $\omega^{\mathcal{A}}_{l}=\frac{|\mathcal{A}|!\left(|\mathcal{S}_{l}|-|\mathcal{A}|-1\right)!}{|\mathcal{S}_{l}|!}$.\\
                Calculate $v_{s_{l}}^{l}=v_{s_{l}}^{l}+\omega^{\mathcal{A}}_{l}\left[p^{D}_{\mathrm{U}\left(\mathcal{A}\cup\{s_{l}\}\right)}-p^{D}_{\mathrm{U}\left(\mathcal{A}\right)}\right]$
            }
        }
    Obtain $s_{l}^{*}=\arg\max_{s_{l}}v_{s_{l}}^{l}$\\
    \eIf{$p^{D}_{\mathrm{U}\left({\{s_{l}^{*}\}}\right)}>p_{\mathrm{th}}$}
    {
        Obtain $\mathcal{U}^{*}=\mathcal{U}_{s_{l}^{*}}$, $\bar{\mathcal{S}}=\mathcal{S}_{l}$ and $\bar{S}=s_{l}^{*}$.\\
        break.
    }
    {
        $s_{l}'=\arg\max_{s_{l}\neq s_{l}^{*}}v_{s_{l}}^{l}$\\
        Aggregate the object regions $s_{l}^{*}$ and $s_{l}'$ to form a new object region $s_{l}^{*}$.\\
        Update $\mathcal{S}_{l+1}=\mathcal{S}_{l}\setminus\{s_{l}'\}$.\\
        $l=l+1$.
    }
    }
    \KwOut{$\mathcal{U}^{*}$, $\bar{\mathcal{S}}$, $\bar{S}$}
\end{algorithm}
The detailed process of calculating the shapley value $v_{s}$ is shown in Fig. \ref{Fig:Shapley calculation}, containing the following steps:
\begin{itemize}
    \item For each subset $\mathcal{A} \in \mathcal{A}_{s}$, assume that object regions indexed by $\mathcal{A}$ are compressed with target quality factor $q_{t}$ and protected with channel coding (BER=$\epsilon_{t}$), while the rest are compressed with basic quality factor $q_{b}$ and unprotected (BER=$\epsilon_{c}$). The classification probability for the corresponding reconstructed image $\mathrm{U}(\mathcal{A})$ is denoted by $p^{D}_{\mathrm{U}(\mathcal{A})}$.
    \item Assume that object regions indexed by $s$ and $\mathcal{A}$ are both compressed with target quality factor $q_{t}$ and protected with channel coding (BER=$\epsilon_{t}$), while the rest are compressed with basic quality factor $q_{b}$ and unprotected (BER=$\epsilon_{c}$). The classification probability for the corresponding reconstructed image $\mathrm{U}\left(\mathcal{A}\cup\{s\}\right)$ is denoted as $p^{D}_{\mathrm{U}\left(\mathcal{A}\cup\{s\}\right)}$.
    \item For subset $\mathcal{A}$, the contribution of the object region $s$ is calculated as the improvement in classification probability:  
    $\left[p^{D}_{\mathrm{U}\left(\mathcal{A}\cup\{s\}\right)}-p^{D}_{\mathrm{U}\left(\mathcal{A}\right)}\right]$.
    \item As the likelihood of subset $\mathcal{A}$ occurring is $\omega^{\mathcal{A}}=\left(\frac{|\mathcal{A}|!\left(|\mathcal{S}|-|\mathcal{A}|-1\right)!}{|\mathcal{S}|!}\right)$, the marginal contribution of the object region $s$ is calculated as its contribution times the likelihood $\omega^{\mathcal{A}}$.
    \item The shapley value of the object region $s$ is calculated as the sum of the marginal contributions for all subsets $\mathcal{A}\in\mathcal{A}_{s}$.
\end{itemize}
Each object region's shapley value represents its contribution to classify the class $D$. Based on that, the most important superpixel $\mathcal{U}^{*}$ can be obtained in an iteration process. For each iteration $l$, the shapley value $v_{s_{l}}^{l}$ of the object region $s_{l}$ in current object region index set $\mathcal{S}_{l}$ can be calculated using Eq. (\ref{eq:SHAP}). For object region $s_{l}^{*}$ which has the highest shapley value, if its corresponding classification probability $p^{D}_{\mathrm{U}\left(\{s_{l}^{*}\}\right)}$ exceeds the threshold $p_{\mathrm{th}}$, the iteration ends, generating the most important superpixel $\mathcal{U}^{*}=\mathcal{U}_{s_{l}^{*}}$, current object region index set $\bar{\mathcal{S}}=\mathcal{S}_{l}$, and the index of the most important superpixel $\bar{S}=s_{l}^{*}$ in $\bar{\mathcal{S}}$. Otherwise, the object regions $\left(s_{l}^{*},s_{l}'\right)$ with the top 2 shapley values are aggregated to form a new object region, updating the object region index set as $\mathcal{S}_{l+1}=\mathcal{S}\setminus\{s_{l}'\}$. In this process, $\mathcal{U}^{*}$ can be obtained by aggregating a minimum number of object regions. The whole process is summarized in \textbf{Algorithm 1}.

\subsubsection{Semantic information-based importance ranking}
Once the most important superpixel $\mathcal{U}^{*}$ is extracted, the contributions of the remaining object regions to classify the class $D$ can be quantified using the shapley value, with the assumption that $\mathcal{U}^{*}$ has already been compressed by target quality factor $q_{t}$ and protected by channel coding with target BER $\epsilon_{t}$. This is achieved by our proposed semantic information-based importance ranking method. We obtain the index set of the remaining object regions as $\widetilde{\mathcal{S}}=\bar{\mathcal{S}}\setminus\{\bar{S}\}$. For each remaining object region with index $\widetilde{s}\in\widetilde{\mathcal{S}}$, let $\mathcal{A}_{\widetilde{s}}=\{\mathcal{A}\subseteq\widetilde{\mathcal{S}}|\widetilde{s}\notin\mathcal{A}\}$ denote all subsets of $\widetilde{\mathcal{S}}$ that exclude the index $\widetilde{s}$. Similar to the previous shapley value calculation process, the semantic information-based importance ranking method has the following steps:
\begin{itemize}
    \item For each subset $\mathcal{A}\in\mathcal{A}_{\widetilde{s}}$, assume that the object regions with indexes in the subset $\mathcal{A}$ and the most important superpixel are compressed with target quality factor $q_{t}$ and protected with channel coding (BER=$\epsilon_{t}$) while the rest are compressed with basic quality factor $q_{b}$ and unprotected (BER=$\epsilon_{c}$). The corresponding classification probability is denoted as $p^{D}_{\mathrm{U}\left(\mathcal{A}\cup\{\widetilde{S}\}\right)}$.
    \item Assume the object regions indexed by $\widetilde{s}$ and $\mathcal{A}$, and the most important superpixel are compressed with target quality factor $q_{t}$ and protected with channel coding (BER=$\epsilon_{t}$), while the rest are compressed with basic quality factor $q_{b}$ and unprotected (BER=$\epsilon_{c}$). The corresponding classification probability is denoted as $p^{D}_{\mathrm{U}\left(\mathcal{A}\cup\{\widetilde{S},\widetilde{s}\}\right)}$.
    \item For subset $\mathcal{A}$, the contribution of the object region $\widetilde{s}$ is calculated as the improvement in classification probability $\left[p^{D}_{\mathrm{U}\left(\mathcal{A}\cup\{\widetilde{S},\widetilde{s}\}\right)}-p^{D}_{\mathrm{U}\left(\mathcal{A}\cup\{\widetilde{S}\}\right)}\right]$.
    \item As the likelihood of subset $\mathcal{A}$ occurring is $\widetilde{\omega}^{\mathcal{A}}=\left(\frac{|\mathcal{A}|!\left(|\widetilde{\mathcal{S}}|-|\mathcal{A}|-1\right)!}{|\widetilde{\mathcal{S}}|!}\right)$, the marginal contribution of the object region $\widetilde{s}$ is calculated as its contribution times the likelihood $\widetilde{\omega}^{\mathcal{A}}$.
    \item The shapley value of the object region $\widetilde{s}$ is calculated as the sum of the marginal contributions for all subsets $\mathcal{A}\in\mathcal{A}_{\widetilde{s}}$.
\end{itemize}
\begin{algorithm}[h]
    \caption{Semantic information-based importance ranking}
    \KwIn{$\bar{\mathcal{S}}$, $\bar{S}$, $p_{\mathrm{th}}$, $q_t$. $q_b$, $\epsilon_{c}, \epsilon_{t}$.} 
    Initialize $\widetilde{\mathcal{S}}=\bar{\mathcal{S}}\setminus\{\bar{S}\}$, $\mathcal{U}^{+}=\emptyset$, $\mathcal{U}^{-}=\emptyset$.\\
    Load classification model.\\
    
        \For{$\widetilde{s}\in\widetilde{\mathcal{S}}$}
        {
            $\mathcal{A}_{\widetilde{s}}=\{\mathcal{A}\subseteq\widetilde{\mathcal{S}}|\widetilde{s}\notin\mathcal{A}\}$.\\
            $v_{\widetilde{s}}=0$\\
            \For{$\mathcal{A}\in\mathcal{A}_{\widetilde{s}}$}
            {
                Obtain $p^{D}_{\mathrm{U}\left(\mathcal{A}\cup\{\widetilde{S}\}\right)}$ and $p^{D}_{\mathrm{U}\left(\mathcal{A}\cup\{\widetilde{S},\widetilde{s}\}\right)}$ from classification model.\\
                Calculate $\widetilde{\omega}^{\mathcal{A}}=\frac{|\mathcal{A}|!\left(|\widetilde{\mathcal{S}}|-|\mathcal{A}|-1\right)!}{|\widetilde{\mathcal{S}}|!}$.\\
                Calculate $v_{\widetilde{s}}=v_{\widetilde{s}}+\widetilde{\omega}^{\mathcal{A}}\left[p^{D}_{\mathrm{U}\left(\mathcal{A}\cup\{\widetilde{S},\widetilde{s}\}\right)}-p^{D}_{\mathrm{U}\left(\mathcal{A}\cup\{\widetilde{S}\}\right)}\right]$
            }
            \eIf{$v_{\widetilde{s}}\geq0$}
            {
            $\mathcal{U}^{+}=\mathcal{U}^{+}\cup\mathcal{U}_{\widetilde{s}}$.\\
            }
            {
            $\mathcal{U}^{-}=\mathcal{U}^{-}\cup\mathcal{U}_{\widetilde{s}}$.\\
            }
        }
    
    \KwOut{$\mathcal{U}^{+}$, $\mathcal{U}^{-}$.}
\end{algorithm}
Upon completing the above process, the remaining object regions can be classified into positive superpixels $\mathcal{U}^{+}$ and negative superpixels $\mathcal{U}^{-}$. $\mathcal{U}^{+}$ consists of the remaining object regions with positive shapley values, indicating that these regions could improve the classification probability with target compression quality factor and channel coding protection. $\mathcal{U}^{-}$ includes the remaining object regions with negative shapley values, signifying that applying target compression quality factor and channel coding protection to these regions degrades the classification probability. The whole process is summarized in \textbf{Algorithm 2}.

\par In practice, the need to compute Shapley values does not prevent real-time deployment. First, we could work on a limited number of object regions, which already reduces complexity. Moreover, shapley values can be obtained via standard sampling-based or approximate methods that only aim to produce a reliable ranking of region importance. Finally, these importance maps can be precomputed or updated at the transmitter on a slower time scale (e.g., per scene).

\subsection{Semantic Coding Design}
\par Given the basic compression quality factor $q_{b}$ and channel BER $\epsilon_{c}$, applying a higher target compression quality factor $q_{t}$ and channel coding protection (achieve BER $\epsilon_{t}$) to the most important superpixel $\mathcal{U}^{*}$ is sufficient to guarantee $p^{\mathrm{D}}_{\hat{\boldsymbol{u}}}>p_{\mathrm{th}}$. Thus, to minimize the overall transmission cost, our semantic coding design focuses on two aspects: semantic source coding design and semantic channel coding design.

\subsubsection{Semantic Source Coding Design}
Semantic source coding design aims to compress the most important superpixel $\mathcal{U}^{*}$ at the target quality factor $q_{t}$ and compress the remaining image regions $\mathcal{U}^{\mathrm{R}}$ at the basic quality factor $q_{b}$. To achieve this, our proposed G-JSSCC framework uses the standard source coding schemes that are compatible with current standards, such as JPEG and JPEG 2000, where compression quality factor can be explicitly controlled.
\begin{figure}[htbp]
    \centering
    \includegraphics[width=\linewidth]{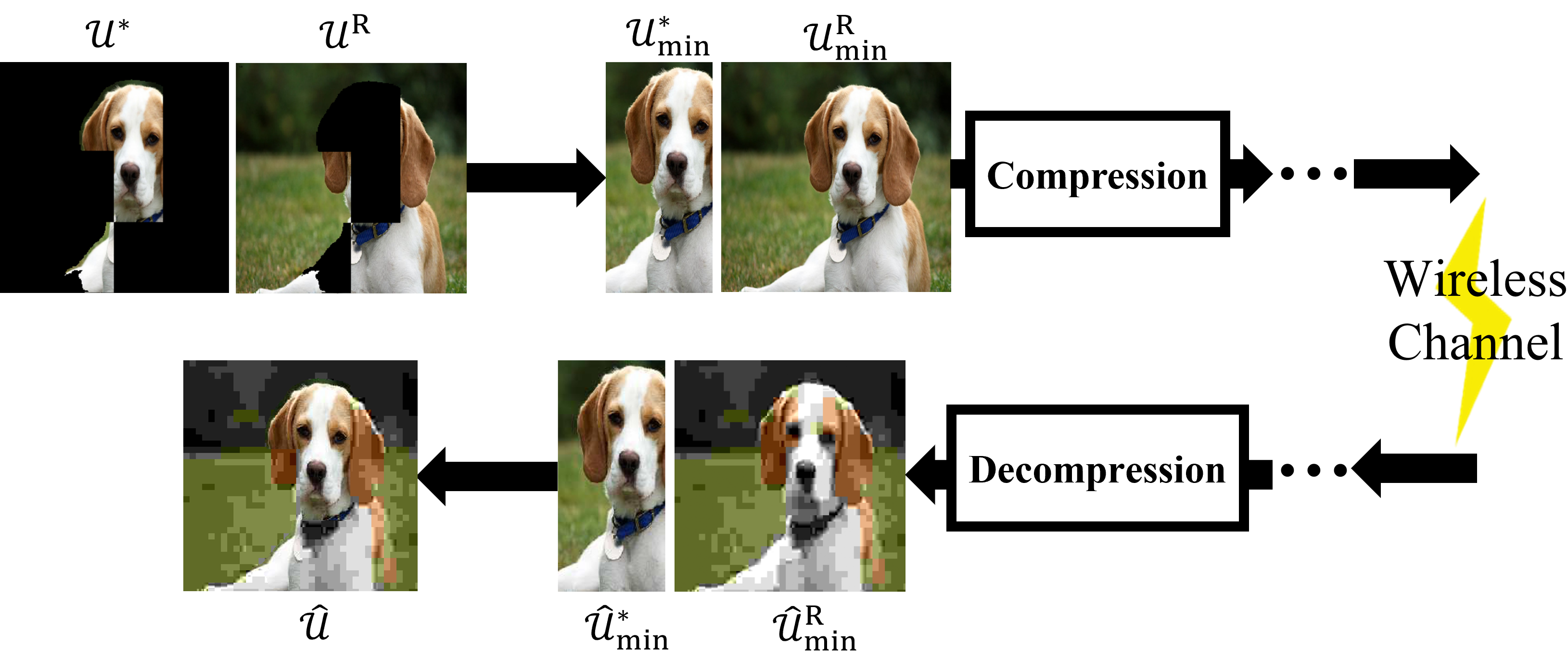}
    \caption{Semantic source coding design.}
    \label{Fig:Seman_Source_Coding}
\end{figure}
\par The semantic source coding design is shown in Fig. \ref{Fig:Seman_Source_Coding}. At the transmitter, JPEG and JPEG 2000 require rectangular input image areas, whereas superpixels are typically irregular in shape. To accommodate this, we need to obtain the rectangular area containing the image regions that need to be compressed. For a given compression quality factor, a larger input rectangular region yields a higher number of compressed bits. To reduce transmission cost by generating as few compressed bits as possible, we first identify the minimum rectangular areas $\mathcal{U}^{*}_{\mathrm{min}}$ and $\mathcal{U^{\mathrm{R}}_{\mathrm{min}}}$, which enclose the most important superpixel $\mathcal{U}^{*}$ and other remaining image regions $\mathcal{U^{\mathrm{R}}}$, respectively. These rectangular areas are then compressed separately, with $\mathcal{U}^{*}_{\mathrm{min}}$ compressed to reach the target quality factor $q_{t}$ and $\mathcal{U}^{\mathrm{R}}_{\mathrm{min}}$ compressed to reach the basic quality factor $q_{b}$. At the receiver, the two rectangular areas are independently decompressed. The decompressed rectangular area $\hat{\mathcal{U}}_{\mathrm{min}}^{*}$, corresponding to $\mathcal{U}^{*}_{\mathrm{min}}$, retains only the content aligned with $\mathcal{U}^{*}$. Similarly, the decompressed rectangular area $\hat{\mathcal{U}}_{\mathrm{min}}^{\mathrm{R}}$, corresponding to $\mathcal{U}^{\mathrm{R}}_{\mathrm{min}}$, preserves only the content aligned with $\mathcal{U}^{\mathrm{R}}$. These preserved components are then combined to reconstruct the final image $\hat{\mathcal{U}}$ for classification.

\subsubsection{Semantic Channel Coding Design}
Semantic channel coding design aims to only protect the bit stream $\boldsymbol{b}^*$ of the most important superpixel $\mathcal{U}^{*}$ to minimize the total transmitted bits. For $\boldsymbol{b}^*$ with the number of bits $K^*$, the aim of channel coding is to reduce the BER introduced by the channel $\epsilon_{c}$ to the target BER $\epsilon_{t}$. As a result, the problem is formed as designing a coding scheme operating at an SNR of $\rho = (Q^{-1}(\epsilon_c))^2$ per real dimension according to \eqref{eq:BPSK_BER} to achieve a BER no larger than $\epsilon_{t}$. Next, we use the normal approximation (NA) bound for BPSK signaling to guide our code design to achieve the maximum coding rate or the minimum codeword length, which in turn maximizes $R$. To use the NA bound, we need to obtain the target frame error rate (FER). Since $\text{BER} \leq \text{FER}$, we let $\text{FER} \leq \epsilon_{t}$ to guarantee that the decoder's output BER is no larger than $\epsilon_{t}$. Using the NA bound \cite{5452208}, the number of bits $K^{*}$ satisfies
\begin{align}\label{eq:NA}
K^* \leq I(X;Y)N^*-\sqrt{V(X;Y)N^*}Q^{-1}(\epsilon_t)+\frac{1}{2}\log_2 N^*,
\end{align}
where the mutual information is
\begin{align}
I(X;Y) %=& \mathbb{E}[i(X;Y)|h],
=  \frac{1}{\sqrt{2\pi}}\int^{\infty}_{-\infty}e^{-\frac{z^2}{2}}\left(1 - \log_2(1+e^{-2\rho-2z\sqrt{\rho}}) \right)dz,
\end{align}
and the channel dispersion is
\begin{align}
V(X;Y)
=\frac{1}{\sqrt{2\pi}}\int^{\infty}_{-\infty}e^{-\frac{z^2}{2}}\left(1 - \log_2(1+e^{-2\rho-2z\sqrt{\rho}}) \right)^2dz,
%& - (I(X;Y))^2
\end{align} 
Given $K^*$, $\epsilon_t$, and $\epsilon_c$, the required minimum codeword length $N^{*}$ can be determined from \eqref{eq:NA}. Finally, the problem reduces to designing a channel code with codeword length $N^*$ and code rate $\frac{K^*}{N^*}$.

\section{Simulation Results}
In this section, we evaluate the performance of our proposed G-JSSCC framework which only employs target compression quality factor and channel coding protection to the most important superpixel $\mathcal{U}^{*}$, compared to three schemes:
\begin{itemize}
    \item $\mathcal{U}^{*}$ and $\mathcal{U}^{+}$: Only apply target compression quality factor and channel coding protection to the most important $\mathcal{U}^{*}$ and positive $\mathcal{U}^{+}$ superpixels.
    \item $\mathcal{U}^{*}$ and $\mathcal{U}^{-}$: Only apply target compression quality factor and channel coding protection to the most important $\mathcal{U}^{*}$ and negative $\mathcal{U}^{-}$ superpixels.
    \item $\mathcal{U}$: Apply target compression quality factor and channel coding protection to the full image $\mathcal{U}$.
\end{itemize}
To see how BER after channel coding would impact classification, we apply GSC-PCCs to achieve different rates in protecting only $\mathcal{U}^{*}$, while other three schemes use ideal i.i.d. coding over BPSK signaling for simplicity. 

\par We employ the InceptionV3 model, pre-trained on the ImageNet dataset, which outputs a probability distribution over $C=1000$ classes. A U-Net model trained on the Oxford-IIIT Pet Dataset is used for object–background segmentation. The test image\footnote{We emphasize that our proposed framework is applicable to any image. However, since the classification probability $p^D_{\hat{\mathcal{U}}}$ can vary significantly across images within the same category (e.g., 0.2 for one image and 0.8 for another), averaging results over multiple images may obscure the impacts of semantic importance on performance. Therefore, to illustrate the behavior of our scheme and highlight how different semantically important regions influence classification, we showcase results using a single test image, which is randomly selected from the Oxford-IIIT Pet Dataset labeled “beagle”.} is selected from Oxford-IIIT Pet Dataset, labeled as ``beagle", with the threshold $p_{\mathrm{th}}=0.7$. We run our simulations for both high and low channel BER cases. 

\subsection{Poor Channel Quality Case}
\begin{figure}[htbp]
    \centering
    \includegraphics[width=\linewidth]{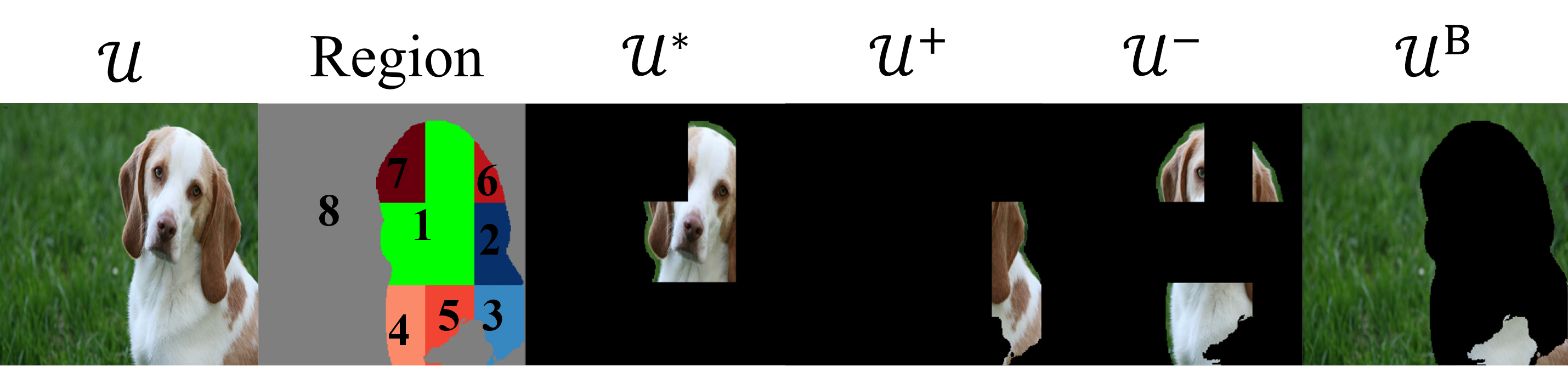}
    \caption{Segmentation after semantic information extraction.}
    \label{Fig:Segmentaion Results}
\end{figure}
 \begin{figure}[htbp]
    \centering
    \subfigure[Codeword length of different schemes when $\epsilon_{t}=10^{-1}$.]{
    \includegraphics[width=0.8\linewidth]{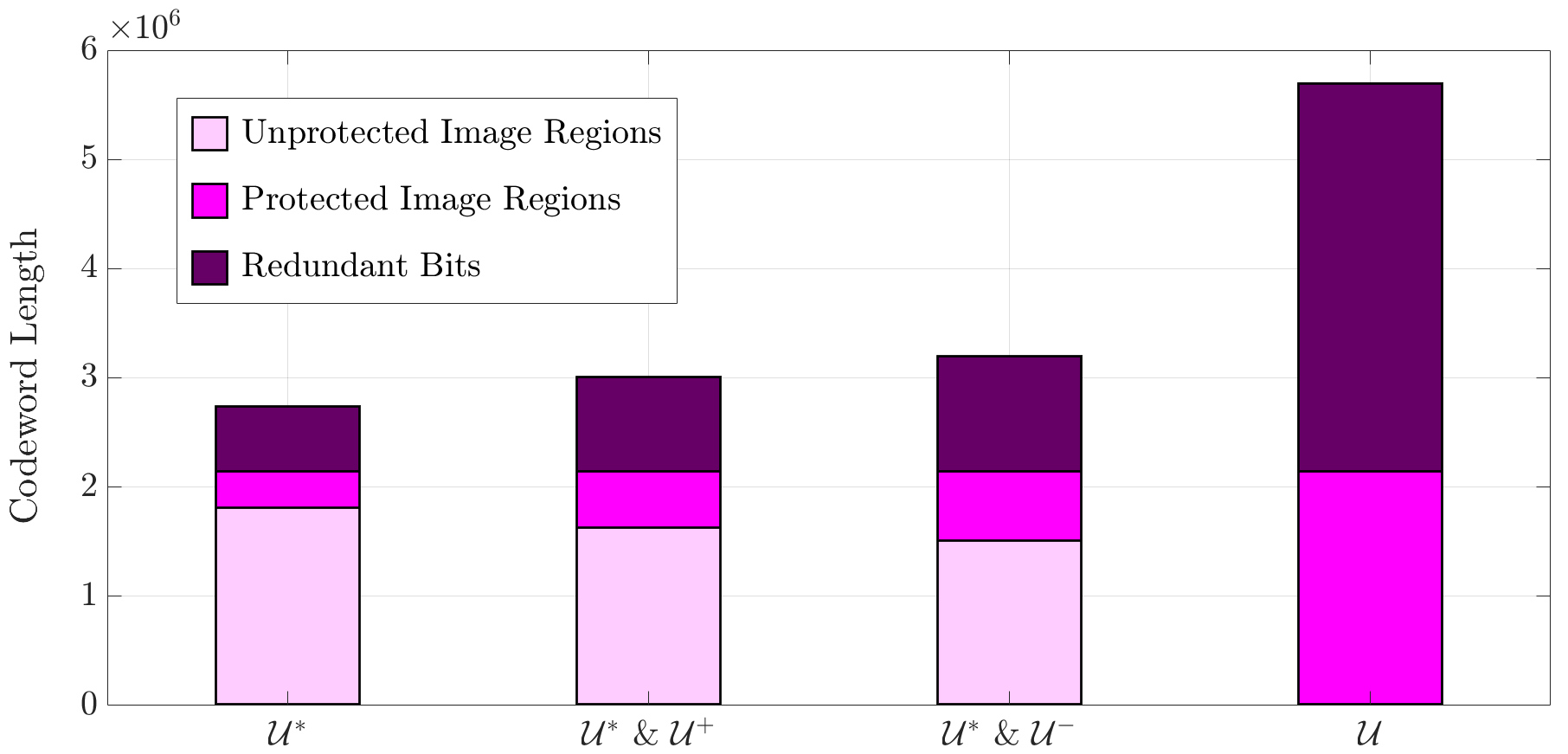}
    }
    \subfigure[Increased redundant bits of different schemes among various $\epsilon_{t}$ compared to $\epsilon_{t}=10^{-1}$.]{
    \includegraphics[width=0.8\linewidth]{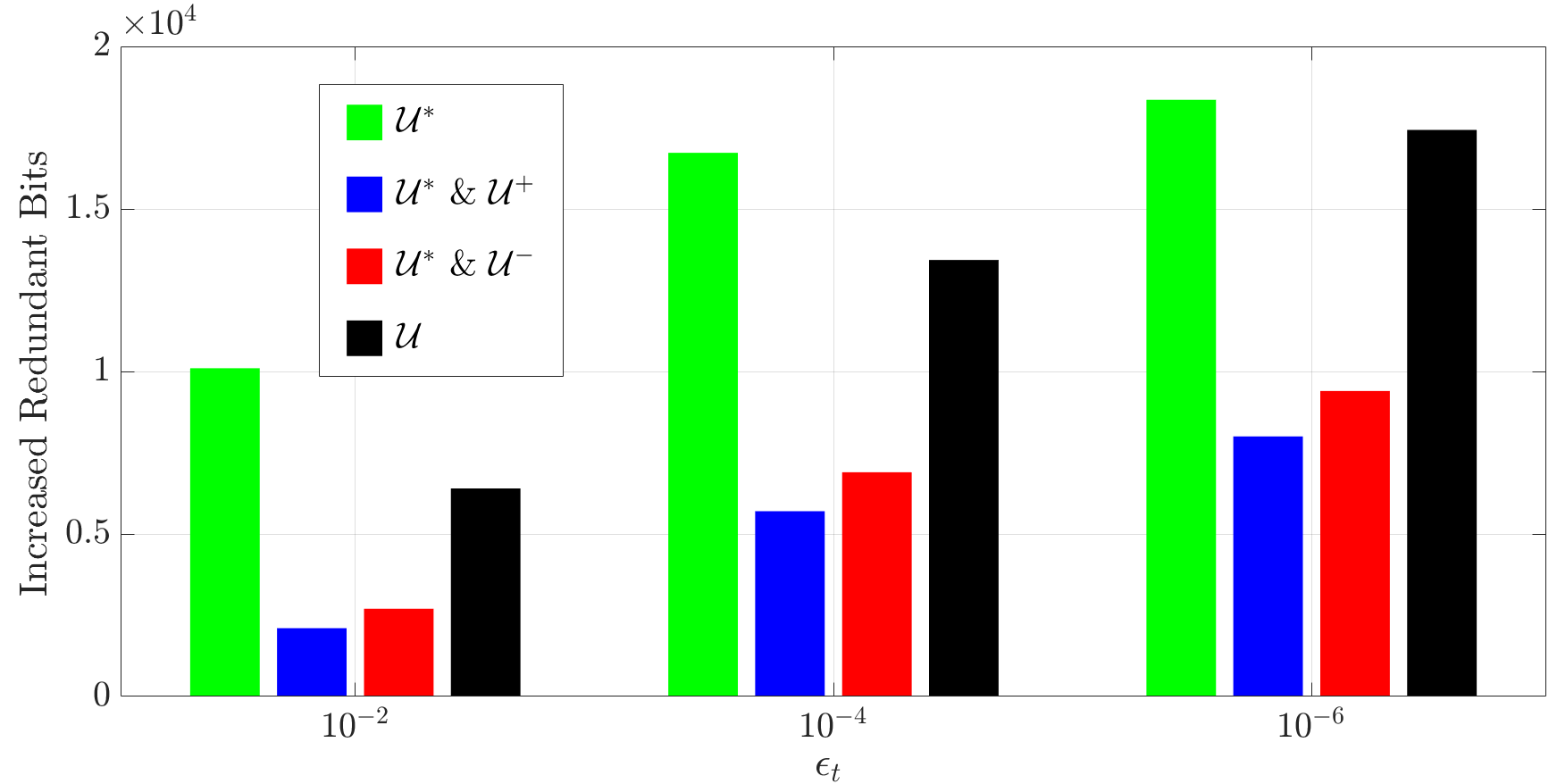}
    }
    \caption{Codeword length after semantic channel coding.}
    \label{Fig:Code Results}
\end{figure}
\begin{figure}[htbp]
    \centering
    \subfigure[$p^{D}_{\hat{\mathcal{U}}}$ comparisons among $\epsilon_{t}$.]{
    \includegraphics[width=0.7\linewidth]{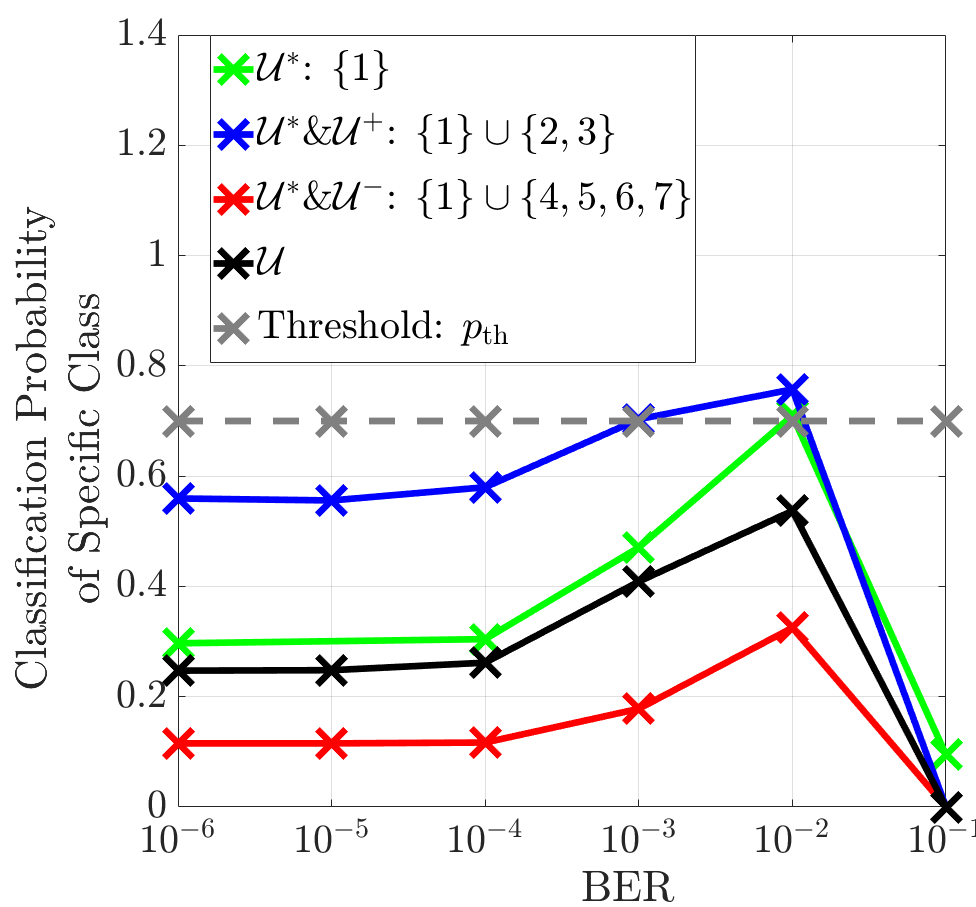}
    }
    \subfigure[$e^{D}_{\hat{\mathcal{U}}}$ comparisons among $\epsilon_{t}$.]{
    \includegraphics[width=0.7\linewidth]{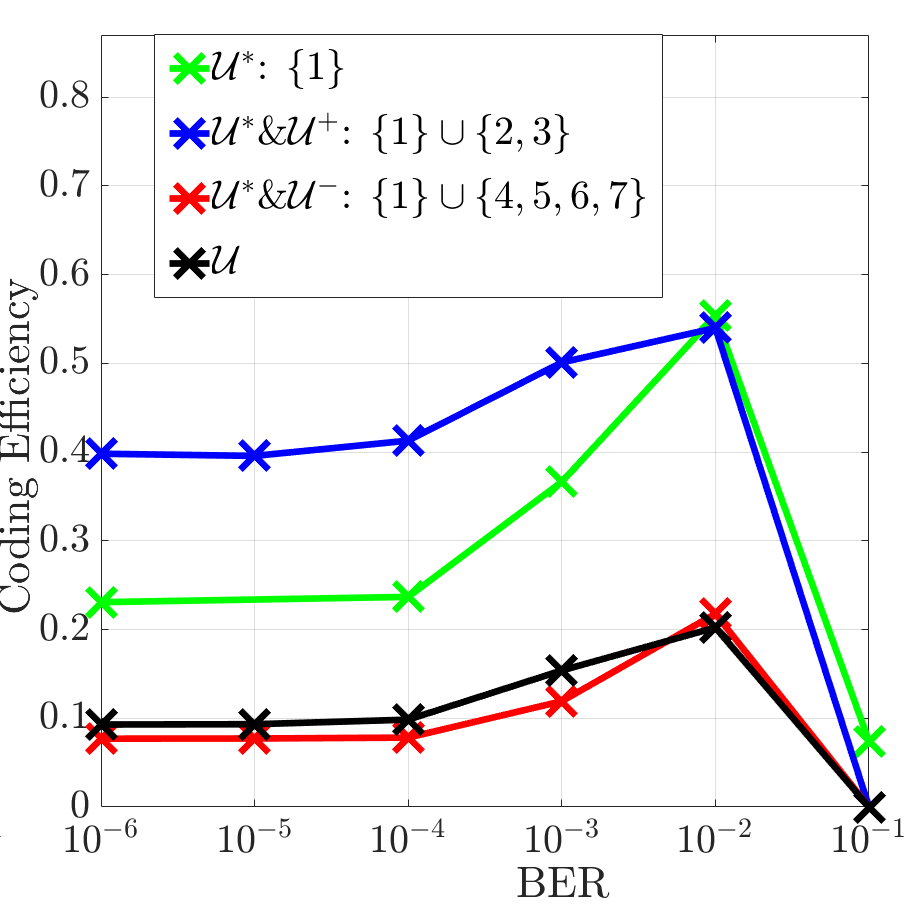}
    }
    \caption{Performance comparisons.}
    \label{Fig:Performance}
\end{figure}
With poor channel quality, we consider $|h|^{2}=0.7$ with the corresponding wireless channel BER as $\epsilon_{c}\approx0.2014$, the target BER after channel coding as $\epsilon_{t}=10^{-2}$, and the threshold $p_{\mathrm{th}}$ to $0.7$. As standard image compression formats such as JPEG and JPEG 2000 utilize entropy coding that is highly sensitive to bit errors, even minor errors in the header or entropy-coded part of the compressed bit stream can cause complete decoding failure over noisy wireless channels \cite{JPEG2000_limit}. Thus, in this case, we only focus on channel coding.

\par Fig. \ref{Fig:Segmentaion Results} plots the segmentation results after semantic information extraction, where the gray, green, blue, and red parts in the segmentation represent the background $\mathcal{U}^{\mathrm{B}}$, most important superpixel $\mathcal{U}^{*}$, positive superpixels $\mathcal{U}^{+}$, and negative superpixels $\mathcal{U}^{-}$. For $\mathcal{U}^{+}$ or $\mathcal{U}^{-}$, a deeper color intensity represents a higher/lower shapley value, indicating a stronger positive/negative impact on the classification probability after channel coding. It is observed that $\mathcal{U}^{*}$ captures discriminative features, which are unique to identify the target ``beagle" class, such as the facial region. $\mathcal{U}^{+}$ contain additional class-relevant features, such as the distinctive ear shape. Conversely, $\mathcal{U}^{-}$ often represent common features shared across multiple dog breeds, such as ear color, which may introduce ambiguity and thereby degrade the classification performance.

\par Based on the segmentation results in Fig. \ref{Fig:Segmentaion Results}, Fig. \ref{Fig:Code Results} plots the codeword length comparisons on various target BER $\epsilon_{t}$. Fig. \ref{Fig:Code Results} (a) presents codeword lengths of the four schemes under $\epsilon_{t}=10^{-1}$. Clearly, protecting only the most important superpixel $\mathcal{U}^{*}$ results in the shortest total codeword length, while full-image protection yields the longest. This demonstrates the efficiency of our proposed G-JSSCC framework in minimizing the amount of data requiring protection, thereby significantly reducing the overhead caused by redundant bits, especially compared to full-image coding. Fig. \ref{Fig:Code Results} (b) plots the increase in the number of redundant bits for the four schemes across various target BER after channel coding $\epsilon_{t}$ compared to $\epsilon_{t} = 10^{-1}$. We can observe that lowering $\epsilon_t$ leads to increased redundant bits for stronger protection. It is also noticed that, only protecting $\mathcal{U}^{*}$ exhibits the highest overall increase in redundant bits, as it employs practical channel coding, while other schemes assume ideal i.i.d. coding.
\begin{figure}[htbp]
    \centering
    \subfigure[JPEG.]{
    \includegraphics[width=0.9\linewidth]{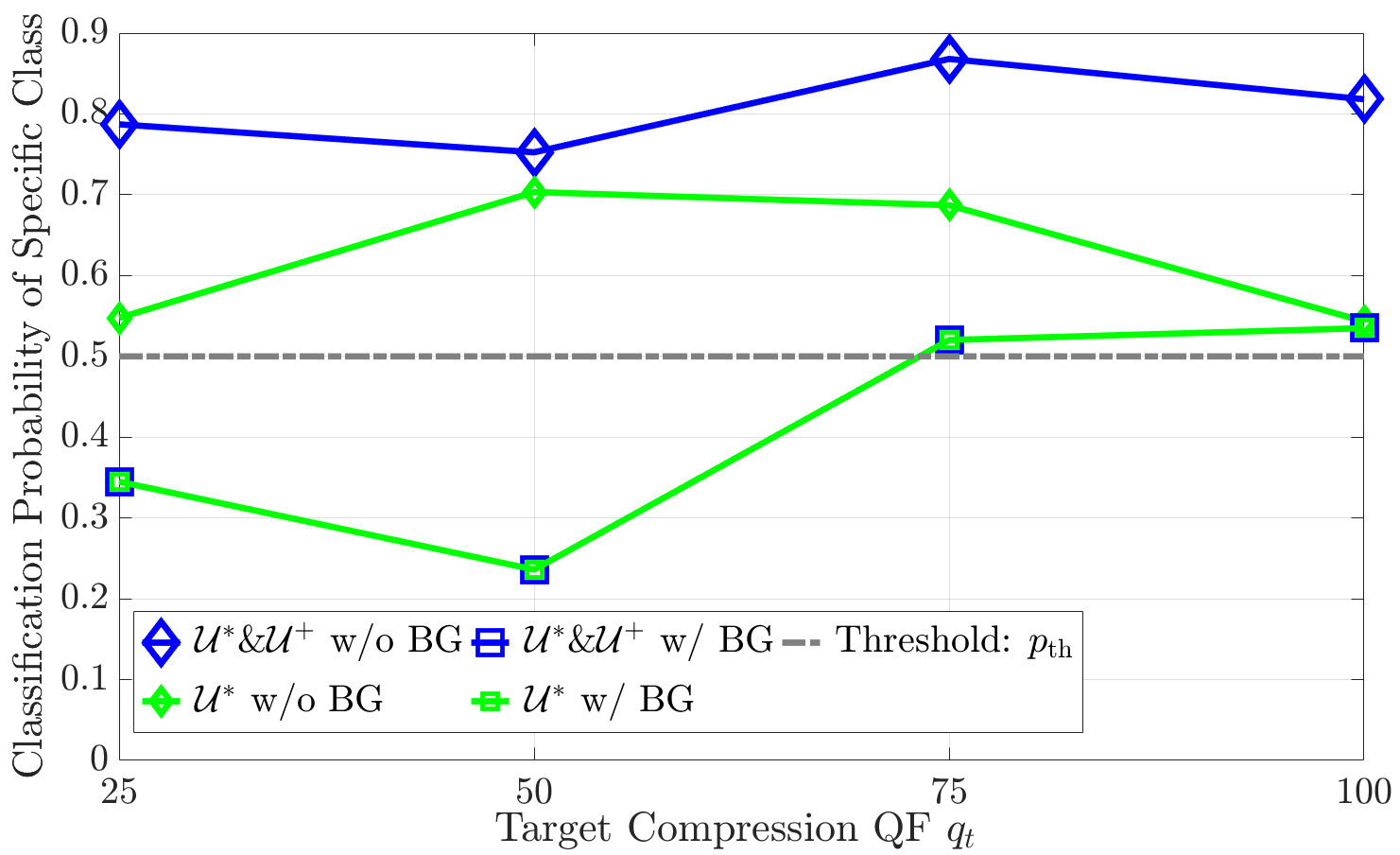}
    }
    \subfigure[JPEG 2000.]{
    \includegraphics[width=0.9\linewidth]{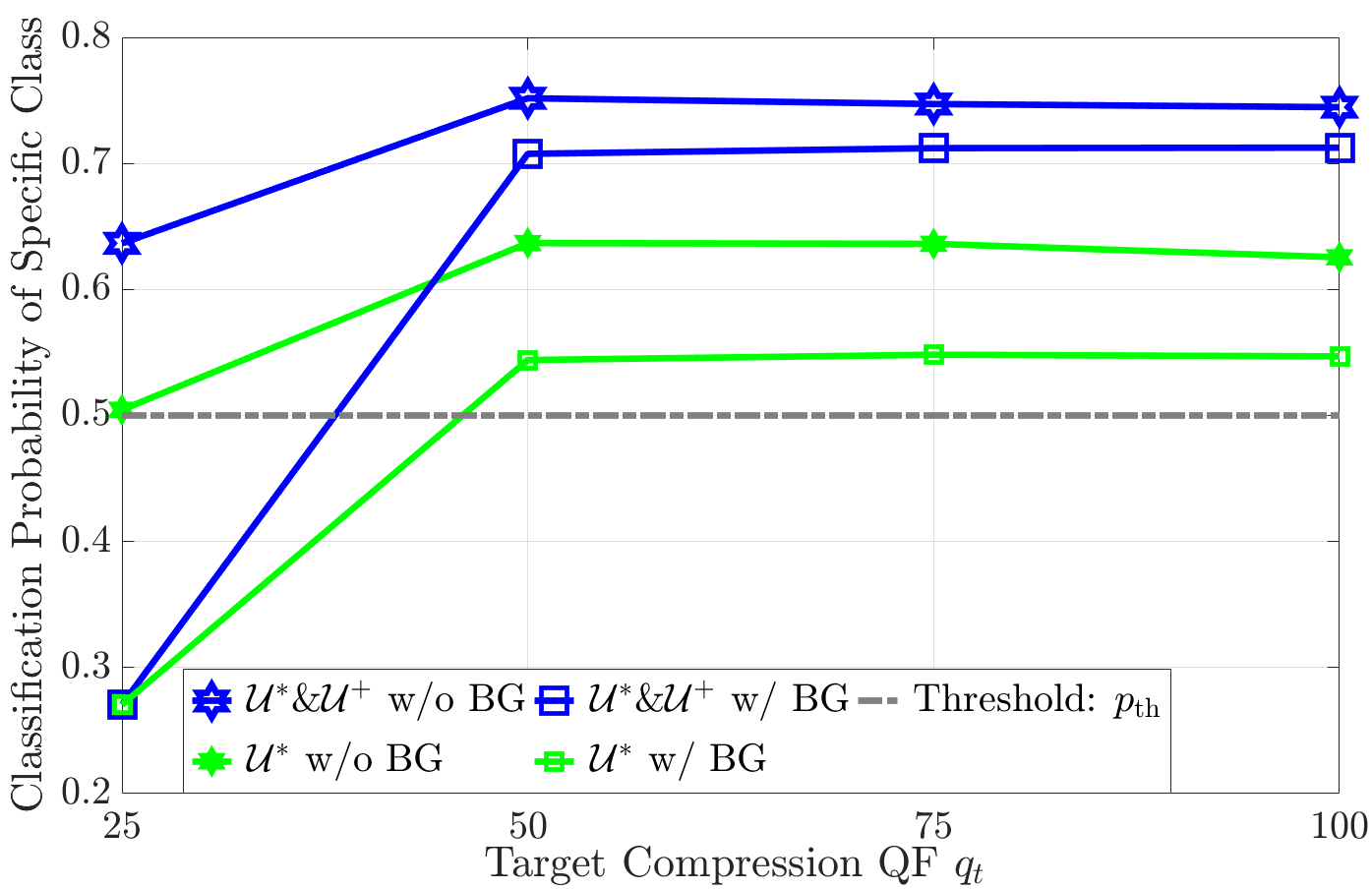}
    }
    \caption{Performance comparisons in source coding perspective with $q_{b}=1$.}
    \label{Fig:Compress_only}
\end{figure}
\par Fig. \ref{Fig:Performance} compares the performances of four schemes when the protected image regions achieve various target BER $\epsilon_t$ by channel coding and the rest experience channel BER $\epsilon_c$, where (a) and (b) plot the classification probability of specific class $p^{D}_{\hat{\mathcal{U}}}$ and coding efficiency $e^{D}_{\hat{\mathcal{U}}}$ comparisons. As $\epsilon_t$ increases, both $p^{D}_{\hat{\mathcal{U}}}$ and $e^{D}_{\hat{\mathcal{U}}}$ first rise and then fall, as the interplay between distinctive features (unique to the class ``beagle") and common features (shared with similar classes). When $\epsilon_t$ increases within a moderate range, distinctive features remain recognizable, while common features become corrupted. This reduces the probability of misclassifications to other classes, thus boosting $p^{D}_{\hat{\mathcal{U}}}$ and $e^{D}_{\hat{\mathcal{U}}}$. However, when $\epsilon_t$ exceeds a tolerable level, even distinctive features are distorted, degrading $p^{D}_{\hat{\mathcal{U}}}$ and $e^{D}_{\hat{\mathcal{U}}}$. Furthermore, protecting only the most important superpixel $\mathcal{U}^{*}$ outperforms full-image protection in $p^{D}_{\hat{\mathcal{U}}}$, as full-image protection protects a huge amount of common features that harm classification. Moreover, when protecting only $\mathcal{U}^{*}$ to achieve a BER near the target BER $\epsilon_{t}=10^{-2}$, $p^{D}_{\hat{\mathcal{U}}}$ exceeds the threshold $p_{\mathrm{th}}$, validating that our proposed G-JSSCC framework can guarantee image classification. Among other three schemes, joint protecting $\mathcal{U}^{*}$ and positive superpixels $\mathcal{U}^{+}$ yields the highest $p^{D}_{\hat{\mathcal{U}}}$, while jointly protecting $\mathcal{U}^{*}$ and negative superpixels $\mathcal{U}^{-}$ yields the lowest $p^{D}_{\hat{\mathcal{U}}}$. This validates the correctness of our proposed semantic information extraction method, which accurately identifies $\mathcal{U}^{+}$ enhancing $p^{D}_{\hat{\mathcal{U}}}$ and $\mathcal{U}^{-}$ degrading $p^{D}_{\hat{\mathcal{U}}}$ after channel coding. In terms of $e^{D}_{\hat{\mathcal{U}}}$, protecting only $\mathcal{U}^{*}$ and jointly protecting $\mathcal{U}^{*}$ and $\mathcal{U}^{+}$ outperform others, because they only protect the distinctive features and class-relevant features to classify the class ``beagle", leading to higher $p^{D}_{\hat{\mathcal{U}}}$ with fewer redundant bits.
\begin{figure}[htbp]
    \centering
    \includegraphics[width=0.9\linewidth]{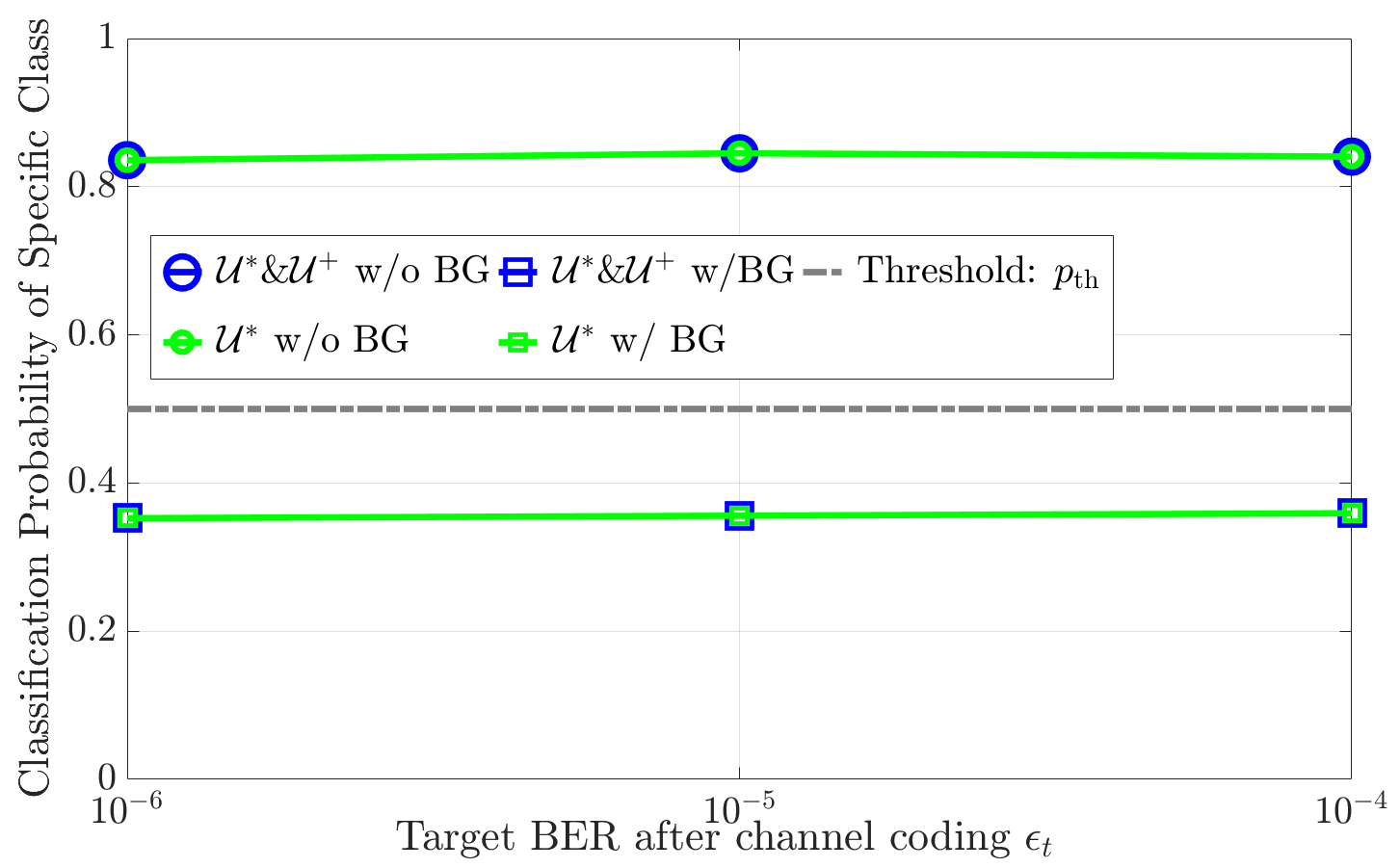}
    \caption{Performance comparisons in channel coding perspective with $\epsilon_{c}=10^{-3}$}
    \label{Fig:Noise_only}
\end{figure}
\begin{figure}[htbp]
    \centering
    \subfigure[Performance comparisons between $\mathcal{U^{*}}\&\mathcal{U}^{+}$ and full image $\mathcal{U}$.]{
    \includegraphics[width=0.8\linewidth]{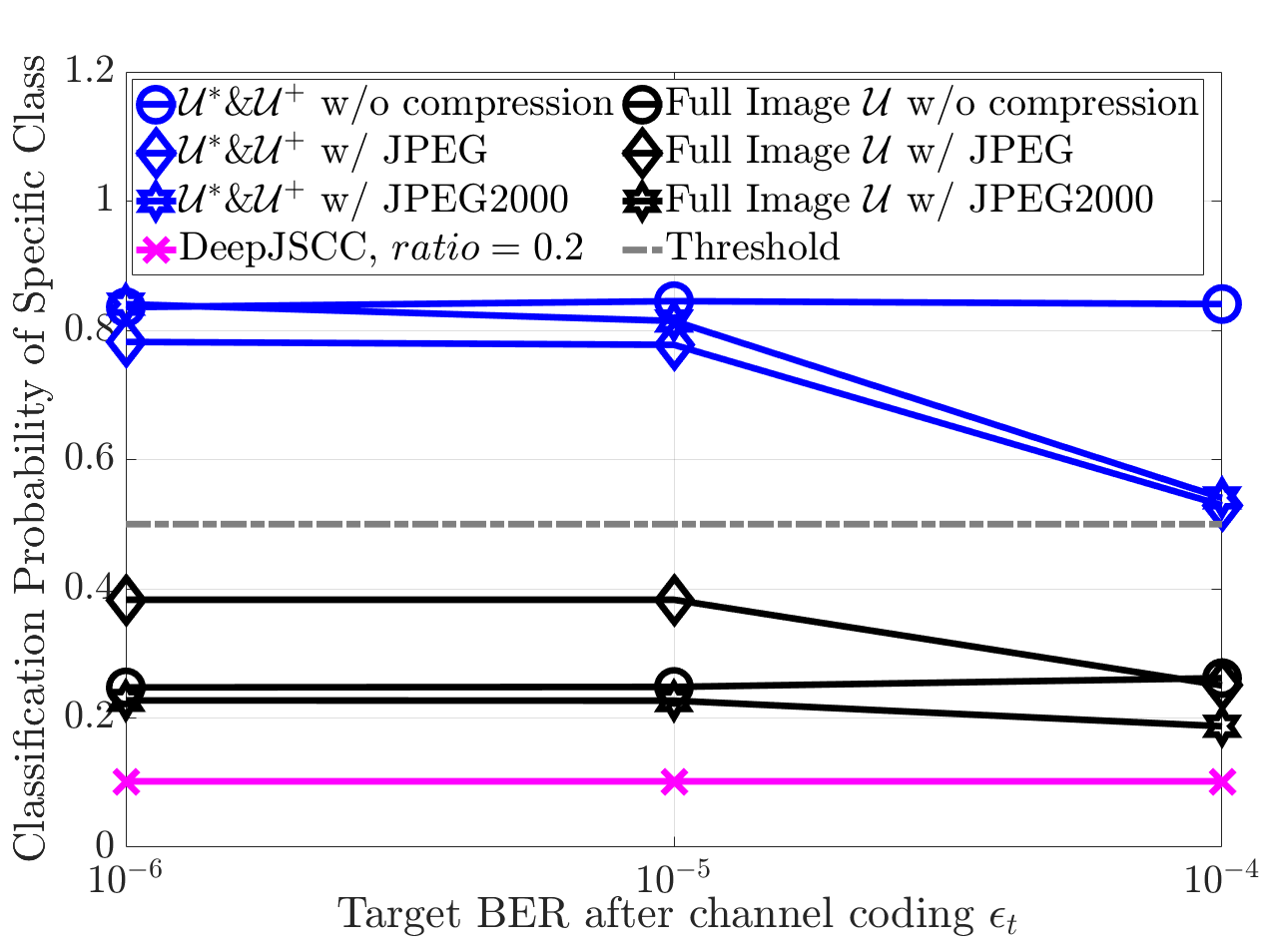}
    }
    \subfigure[Performance comparisons between $\mathcal{U^{*}}$ and $\mathcal{U^{*}}\&\mathcal{U}^{-}$.]{
    \includegraphics[width=0.8\linewidth]{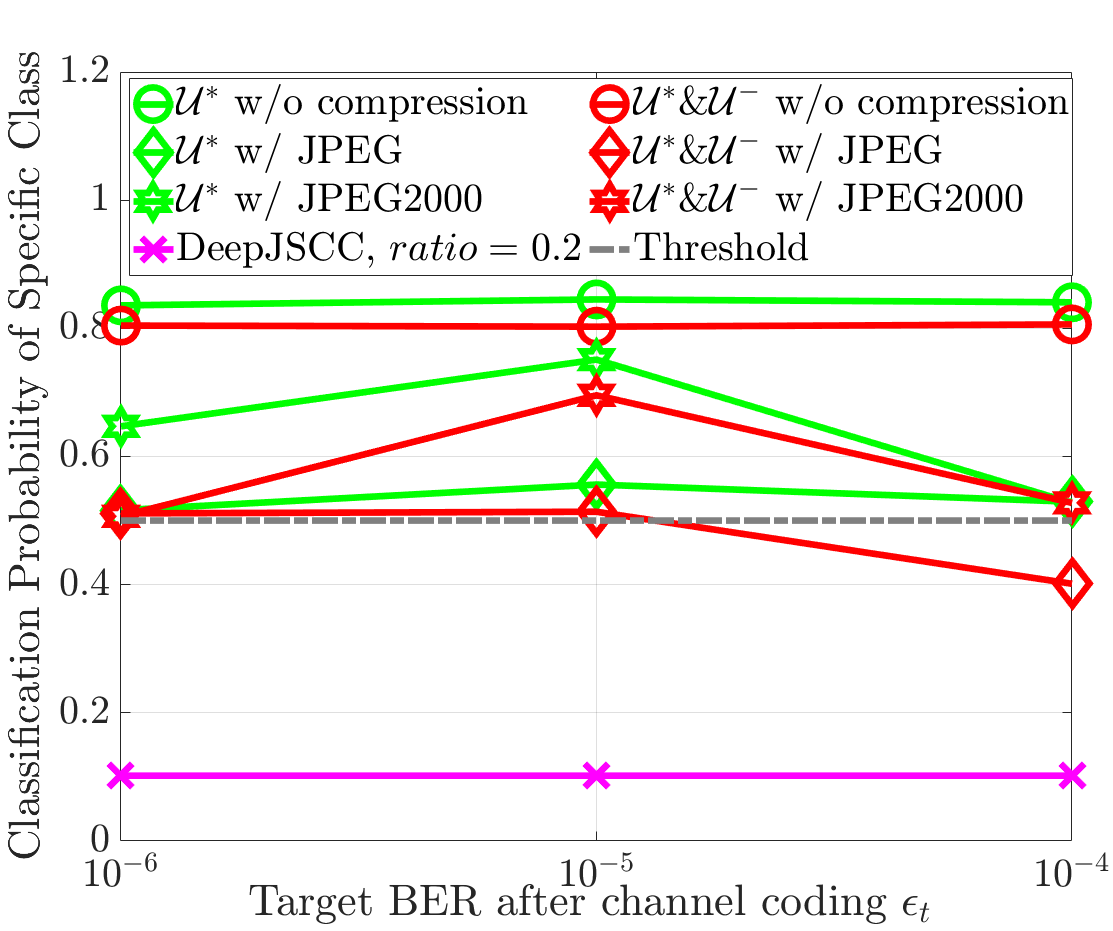}
    }
    \caption{Performance comparisons by jointly considering source and channel coding with $q_{b}=1,q_t=50,\epsilon_c=10^{-3}$ over various $\epsilon_{t}$.}
    \label{Fig:Compress_Noise1}
\end{figure}
\subsection{Good Channel Quality Case}
With good channel quality, we consider $|h|^{2}=9.5495$ with the corresponding wireless channel BER as $\epsilon_{c}\approx10^{-3}$, the target BER after channel coding as $\epsilon_{t}=\{10^{-4},10^{-5},10^{-6}\}$, and the threshold $p_{\mathrm{th}}$ to $0.5$. The basic and target JPEG and JPEG 2000 image compression quality factor are set to be $q_{b}=1$ and $q_{t}=50$, respectively.

\par According to the previous important conclusions \textit{Remark 1} and \textit{Remark 2}, the background part $\mathcal{U}^{\mathrm{B}}$ has no or negative contribution to image classification from both source and channel coding perspectives. Therefore, it is reasonable to consider omitting the transmission of $\mathcal{U}^{B}$. To verify this, we conduct preliminary simulations to compare classification performance with (w/) and without (w/o) background transmission. Since the classification model requires a complete image input, when background is not transmitted, the receiver replaces $\mathcal{U}^{\mathrm{B}}$ with random values drawn from a uniform distribution $U\sim\left(0,255\right)$.

\par Fig. \ref{Fig:Compress_only} plots classification probabilities versus various target compression quality factor $q_{t}$ for JPEG and JPEG 2000 standards, with a fixed basic compression quality factor $q_b=1$, under cases with (square markers) and without (diamond and star markers) background transmission. For both JPEG of Fig. \ref{Fig:Compress_only} (a) and JPEG 2000 of Fig. \ref{Fig:Compress_only} (b), applying the target compression quality factor $q_t$ only to the most important superpixel $\mathcal{U}^{*}$ (green) or to both the most important and positive superpixels $\left(\mathcal{U}^{*}\&\mathcal{U}^{+}\right)$ (blue) without background transmission achieves better classification performance than the corresponding cases with background transmission. This confirms that excluding the background transmission is reasonable from the source coding perspective.

\par Fig. \ref{Fig:Noise_only} plots classification probabilities versus various target BER after channel coding $\epsilon_{t}$ with the channel BER $\epsilon_c=10^{-3}$, under cases with (square markers) and without (round markers) background transmission. Similarly to the phenomena in Fig. \ref{Fig:Compress_only}, applying channel coding only to the most important superpixel $\mathcal{U}^{*}$ (green) or to both the most important and positive superpixels $\left(\mathcal{U}^{*}\&\mathcal{U}^{+}\right)$ (blue) without background transmission achieves better classification performance than the corresponding cases with background transmission. This confirms that excluding the background transmission is also reasonable from the channel coding perspective. Meanwhile, as the channel BER $\epsilon_{c}$ is sufficiently low, the wireless transmission has minimal impact on the quality of the reconstructed image. As a result, the effect of channel coding becomes subtle, leading to only minor variations in the classification probability across different target BER $\epsilon_{t}$.

\par Based on the previous findings in Fig. \ref{Fig:Compress_only} and Fig. \ref{Fig:Noise_only}, subsequent simulations are conducted without background transmission by jointly considering source and channel coding. Fig. \ref{Fig:Compress_Noise1} plots the classification probability $p^{D}_{\hat{\mathcal{U}}}$ comparisons between the four schemes and DeepJSCC versus varying target BER after channel coding $\epsilon_{t}$ with parameter settings $\left(q_b,q_t,\epsilon_c\right)=\left(1,50,10^{-3}\right)$. The four schemes are evaluated under two scenarios: with standard image compression (JPEG and JPEG 2000) and without compression. As shown in Fig. \ref{Fig:Compress_Noise1} (a), applying target compression quality factor and channel coding protection to the most important and positive superpixels $\left(\mathcal{U}^{*}\&\mathcal{U}^{+}\right)$ significantly outperforms those with the full image $\mathcal{U}$ transmission and the DeepJSCC. Moreover, as $\epsilon_t$ increases to $10^{-4}$, the classification probability $p^{D}_{\hat{\mathcal{U}}}$ of using JPEG or JPEG 2000 drops sharply, while the uncompressed case remains stable. This degradation is attributed to the limited error resilience of JPEG-based standards, where even a few bit errors can substantially distort image quality, as also illustrated in Fig. \ref{Fig:Compress_Noise1} (b). Notably, only applying the target compression quality factor and channel coding protection to $\mathcal{U}^{*}$ and $\left(\mathcal{U}^{*}\&\mathcal{U}^{+}\right)$ achieves the overall highest classification probability $p^{D}_{\hat{\mathcal{U}}}$, remaining above the required threshold. This not only validates the robustness of our proposed semantic information extraction method, which accurately identifies positive superpixels $\mathcal{U}^+$ enhancing $p^{D}_{\hat{\mathcal{U}}}$ and negative superpixels $\mathcal{U}^-$ degrading $p^{D}_{\hat{\mathcal{U}}}$, but also demonstrates the robustness and efficiency of our proposed G-JSSCC framework in guaranteeing image classification under varying channel conditions.
 \begin{figure}[htbp]
    \centering
    \includegraphics[width=\linewidth]{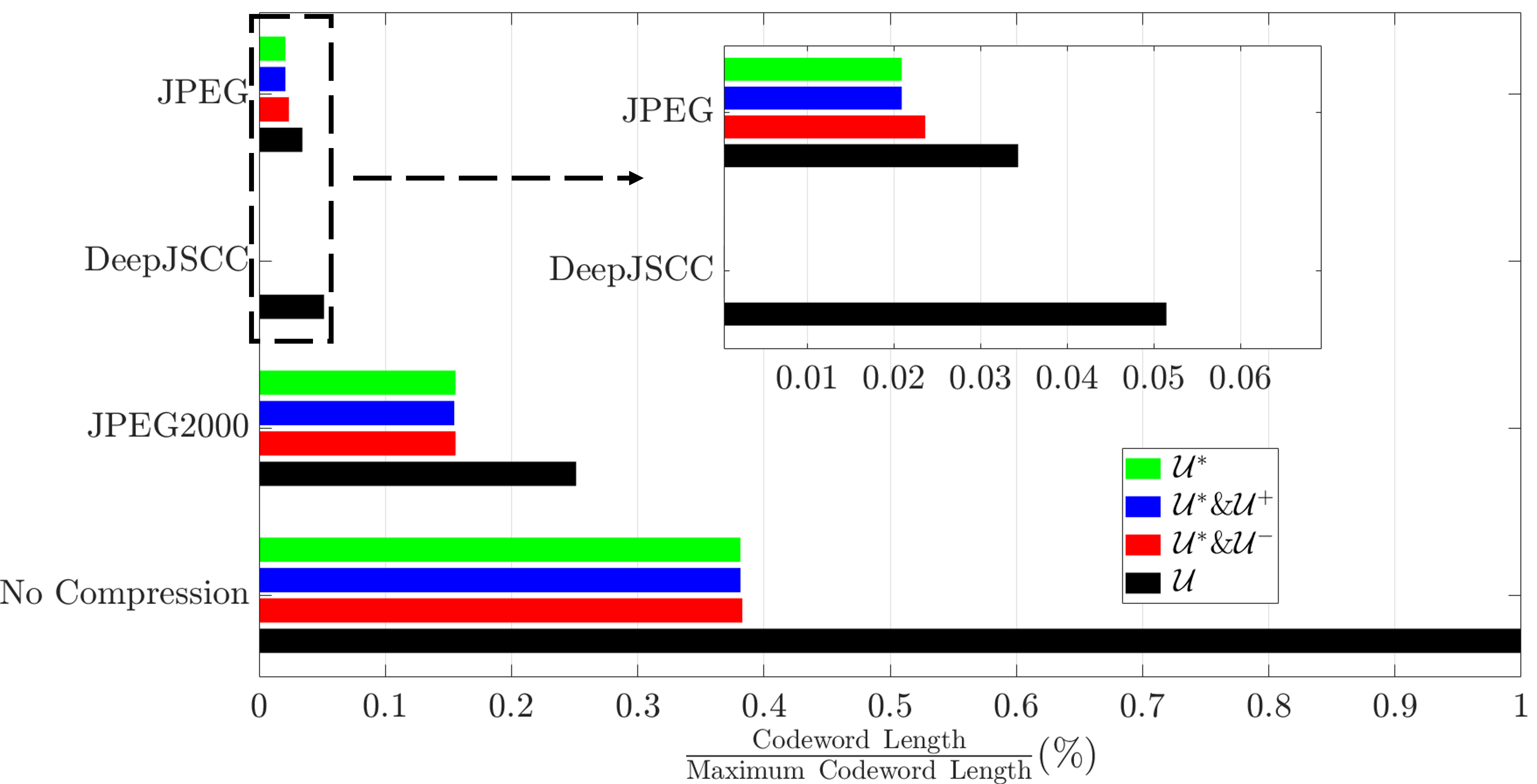}
    \caption{Codeword length comparisons with parameter settings $\left(q_b,q_t,\epsilon_{c},\epsilon_t\right)=\left(1,50,10^{-3},10^{-4}\right)$.}
    \label{Fig:Compress_Noise2}
\end{figure}
\par Fig. \ref{Fig:Compress_Noise2} compares the codeword lengths of DeepJSCC and the four schemes under JPEG, JPEG 2000, and uncompressed cases, with parameters \((q_b, q_t, \epsilon_c, \epsilon_t) = (1, 50, 10^{-3}, 10^{-4})\). Our results clearly demonstrate that applying the target compression quality factor \(q_t\) and channel coding protection exclusively to the most important superpixel \(\mathcal{U}^{*}\) or it with positive superpixels $\left(\mathcal{U}^{*} \& \mathcal{U}^{+}\right)$ yields the shortest codeword lengths across all settings. This validates the efficiency of our proposed G-JSSCC framework in minimizing the overall number of transmitted bits. Moreover, JPEG 2000 generally results in longer codeword lengths than JPEG, due to its emphasis on preserving image quality, whereas JPEG prioritizes compression efficiency at the cost of possible artifacts like blockiness or blurring. Meanwhile, the codeword length of DeepJSCC is a little bit longer than JPEG and shorter than JPEG 2000, implying its strong ability to reduce the transmission cost. As expected, the uncompressed case leads to significantly longer codeword lengths, especially when protecting the full image \(\mathcal{U}\), highlighting the importance of source coding in reducing transmission overhead.
 \begin{figure}[htbp]
    \centering
    \includegraphics[width=0.9\linewidth]{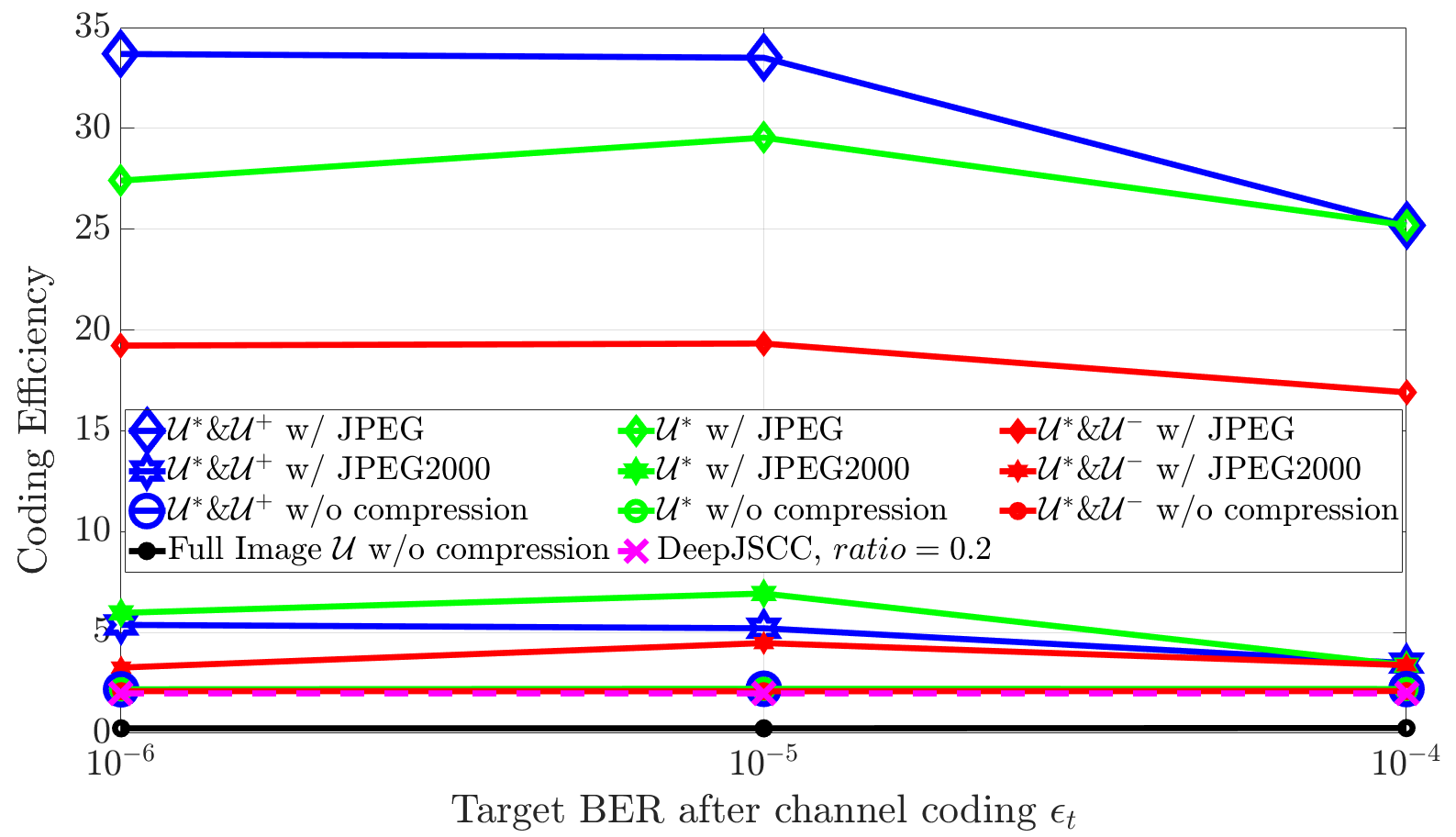}
    \caption{Coding efficiency comparisons.}
    \label{Fig:Compress_Noise3}
\end{figure}
\par Fig. \ref{Fig:Compress_Noise3} compares the coding efficiency for DeepJSCC and the four schemes under varying target BER after channel coding \(\epsilon_t\), across three settings: JPEG, JPEG 2000, and no compression. It is observed that applying the target compression quality factor and channel protection only to the most important superpixel \(\mathcal{U}^{*}\) or that with positive superpixels $\left(\mathcal{U}^{*} \& \mathcal{U}^{+}\right)$ with JPEG consistently achieves the highest efficiency across all values of \(\epsilon_t\). The performance of JPEG 2000 closely follows that of JPEG, particularly for \(\mathcal{U}^{*}\) and $\left(\mathcal{U}^{*} \& \mathcal{U}^{+}\right)$, and significantly outperforms full-image target quality factor compression and channel coding protection. This performance gain is attributed to the fact that applying the target compression quality factor and channel coding protection to \(\mathcal{U}^{*}\) or $\left(\mathcal{U}^{*} \& \mathcal{U}^{+}\right)$ with JPEG or JPEG 2000 can not only maintain high classification probability but also substantially reduces the total number of transmitted bits. Moreover, although DeepJSCC yields a relatively short codeword length, its overall efficiency remains low due to its limited classification probability. Conversely, although applying the target compression quality factor and channel coding protection to \(\mathcal{U}^{*}\) or $\left(\mathcal{U}^{*} \& \mathcal{U}^{+}\right)$ without compression yields a high classification probability, its overall efficiency remains low due to the resulting long codeword length. These results highlight the effectiveness of our proposed G-JSSCC framework in guaranteeing image classification performance while minimizing transmission cost. 

\vspace{-0.2in}
\section{Conclusion}
In this work, we proposed a novel goal-oriented joint semantic source and channel coding (G-JSSCC) framework that guarantees the minimum classification probability while minimizing transmission cost. For a given classification probability threshold, the designed semantic information extraction method identifies and ranks various image regions based on its importance to image classification probability, which is measured by the shapely value from explainable artificial intelligence (AI). Then, our designed semantic source coding and semantic channel coding methods compress and protects the image by applying higher compression quality factor and stronger channel coding protection to more important image region, ensuring classification probability with minimal overall transmission cost. Compared to traditional communication frameworks that uniformly compress and protect the entire image, our proposed G-JSSCC framework improves classification probability by 2.70 times, reduces transmission cost by 38$\%$, and enhances coding efficiency by 5.91 times.

% or


\ifCLASSOPTIONcaptionsoff
  \newpage
\fi



%\bibliographystyle{IEEEtran}
%\begin{thebibliography}{1}
    
    \bibliographystyle{ieeetr}
    \bibliography{mylib}

%\end{thebibliography}


%\newpage


%\vfill


\end{document}